\documentclass[useAMS,usenatbib,usegraphicx]{mn2e}
\usepackage{amsmath}
\usepackage{subfigure}

\newcommand{\subs}[1]{$_{\rm #1}$}

\newcommand{\BE}{\begin{equation}}
\newcommand{\EE}{\end{equation}}
\newcommand{\Msolar}{\mbox{\,$\rm M_{\odot}$}}	      
\newcommand{\kmsn}{km\ s$^{-1}$}
\newcommand{\kmss}{km\ s$^{-1}$ }
\newcommand{\vsinis}{$\!${\em v\,}sin{\em i} }

\def\ga{\mathrel{\hbox{\rlap{\hbox{\lower4pt\hbox{$\sim$}}}\hbox{$>$}}}}
\def\la{\mathrel{\hbox{\rlap{\hbox{\lower4pt\hbox{$\sim$}}}\hbox{$<$}}}}

\title[Detecting planets around young active G and K dwarfs]{Is it possible to detect planets around young active G and K dwarfs?}

\author[S. V. Jeffers et al.]{S.V. Jeffers $^{1}$, J. R.
Barnes$^{2}$, H.R.A. Jones$^{2}$, A. Reiners$^{1}$, D.J. Pinfield$^{2}$, S.C. Marsden$^{3}$\\
$^{1}$ Institut fur Astrophysik, Georg-August-Universitat, Friedrich-Hund-Platz 1, 37077 Goettingen, Germany\\
$^{2}$ Centre for Astrophysics Research, University of Hertfordshire, College Lane, Hatfield, Herts, AL10 9AB, UK\\
$^{3}$ Faculty of Sciences, University of Southern Queensland, Toowoomba, QLD 4350, Australia}

\begin{document}

\date{November 2013}

\pagerange{\pageref{firstpage}--\pageref{lastpage}} \pubyear{2010}

\maketitle

\label{firstpage}

\begin{abstract}
Theoretical predictions suggest that the distribution of planets in very young stars could be very different to that typically observed in Gyr old systems that are the current focus of radial velocity surveys.  However, the detection of planets around young stars is hampered by the increased stellar activity associated with young stars, the signatures of which can bias the detection of planets.  In this paper we place realistic limitations on the possibilities for detecting planets around young active G and K dwarfs.  The models of stellar activity based on tomographic imaging of the G dwarf HD 141943 and the K1 dwarf AB Dor and also include contributions from plage and many small random starspots.  Our results show that the increased stellar activity levels present on young Solar-type stars strongly impacts the detection of Earth-mass and Jupiter mass planets and that the degree of activity jitter is directly correlated with stellar \vsinis.  We also show that for G and K dwarfs, the distribution of activity in individual stars is more important than the differences in induced radial velocities as a function of spectral type.  We conclude that Jupiter mass planets can be detected close-in around fast-rotating young active stars, Neptune-mass planets around moderate rotators and that Super-Earths are only detectable around very slowly rotating stars.  The effects of an increase in stellar activity jitter by observing younger stars can be compensated for by extending the observational base-line to at least 100 epochs.

\end{abstract}


\section{Introduction}



Radial velocity surveys have discovered or detected planetary signatures in 56\% of the confirmed extra-solar planets \footnote{http://exoplanet.eu}. Radial velocity measurements also have the advantage that they can be measured for all stars as opposed to other techniques such as the detection of transit signals.  Most surveys are targeted at detecting planets around older less active stars because signatures of magnetic activity, such as starspots and plage, can at worst mask the presence of a planet and at best distort the absorption line profiles leading to biased  radial velocity measurements.  A recent example is the planet orbiting the young active K5V star BD+20 1790 \citep{Hernan2010}.  However, to fully understand the formation of planets, and how they have evolve, it is also necessary observe planetary systems around young stars. Current observations of planetary systems around young stars are biased as they are mainly observed using direct imaging techniques which favour large and massive planets further out from the host star and are unlikely to be detected using radial velocity techniques even if the host stars were inactive.  

Models of young planetary systems \cite[e.g.][]{Ida2004} have found that planets which migrated into close proximity of their host stars are 1\,-\,10 times more numerous than those at larger orbital radii. A discrepancy of an order of magnitude with observations suggests that a large number of planets that reach close orbital separation are either tidally disrupted \citep{Gu2003} or consumed \citep{Sandquist1998} by their host stars. Once the gas disk dissipates, planets are subject to a stage of evolution that is dominated by gravitational interactions and collisions. During this chaotic era, lasting $\leq 10^8$ yr, many planets may be ejected while the interactions are also able to explain the observed orbital eccentricities of the planets \citep{Chatterjee2008, Ford2008, Juric2008}. One might therefore expect a different distribution of hot Jupiters in young active stellar systems when compared with the typically observed few Gyr old systems with low activity host stars.

It has been well established that magnetic activity declines with age for solar-type stars and that it is correlated with the loss of angular momentum while the star is on the main sequence \citep{Wright2011, Baliunas1995}. Observations of young stars show higher levels of rotation \citep{Soderblom1993} and activity while older stars, with ages comparable to the Sun, show slow rotation levels and lower activity levels \citep{Reiners2009,Covey2011}.  Indeed, increased activity levels are observed in young active stars, using tomographic imaging, and they frequently exhibit large starspots at polar and high latitudes \citep{Strassmeier2009} which is in contrast to the Sun where the spot coverage is confined to equatorial latitudes (i.e. not extending above 35$^\circ$).  The intrinsic stellar variability that leads to additional noise imposed on the radial velocity profile is referred to as jitter and has previously been investigated in detail by \cite{Saar1997} and \cite{Saar1998}, with the stellar activity component being determined by \cite{Wright2005} for the old inactive stars of the Lick and Keck planet searches.  For older stars ($\geq$ 1 Gyr) their results show that there is an empirical relationship between activity, colour and absolute magnitude.

In this paper we aim to place realistic constraints on the detection of planets around active young stars, which show more extreme levels of activity compared to older stars such as the Sun.  To achieve this we use tomographic images of young active stars and model the RV induced jitter caused by the host star's magnetic activity (in the form of starspots and plage).   In Section 2, we summarise the current results of starspot imaging for G and K dwarfs.  In Section 3, we describe the methods to generate the line profiles and we also present our model of planetary parameters.  In Section 4 we determine detection threshold simulations for a range of planetary masses, on a logarithmic scale of planetary orbital radius.  Finally we discuss the probability of detecting planets around active young G and K dwarfs.   

\section[]{Sources of Stellar Jitter}

On the main-sequence, young solar-type stars display magnetic activity signatures that are analogous to those seen on the Sun, in the form of starspots, chromospheric plages, and coronal emission. These are an
order of magnitude or greater than observed on the Sun indicating a powerful magnetic dynamo.   Our current knowledge about starspot distributions on young active stars is based on measurements from tomographic imaging such as Zeeman and Doppler imaging techniques.  The most extensively studied are the G and K dwarfs.

To determine realistic limitations of detecting planets around young active stars it is necessary to use accurate distributions of activity signatures.  In this paper we model the predominant signatures of magnetic activity for a young active G and K dwarf.  The largest starspot regions are taken from starspot distributions reconstructed using tomographic imaging techniques.  Additional activity signatures can also impact the star/planet radial velocity profile.  Examples include bright plage regions and the presence of many unresolved starspots that are below the resolution capability of stellar tomographic techniques.  

\subsection{Starspots on G and K dwarfs}

\begin{figure*} 
\begin{center}
\includegraphics{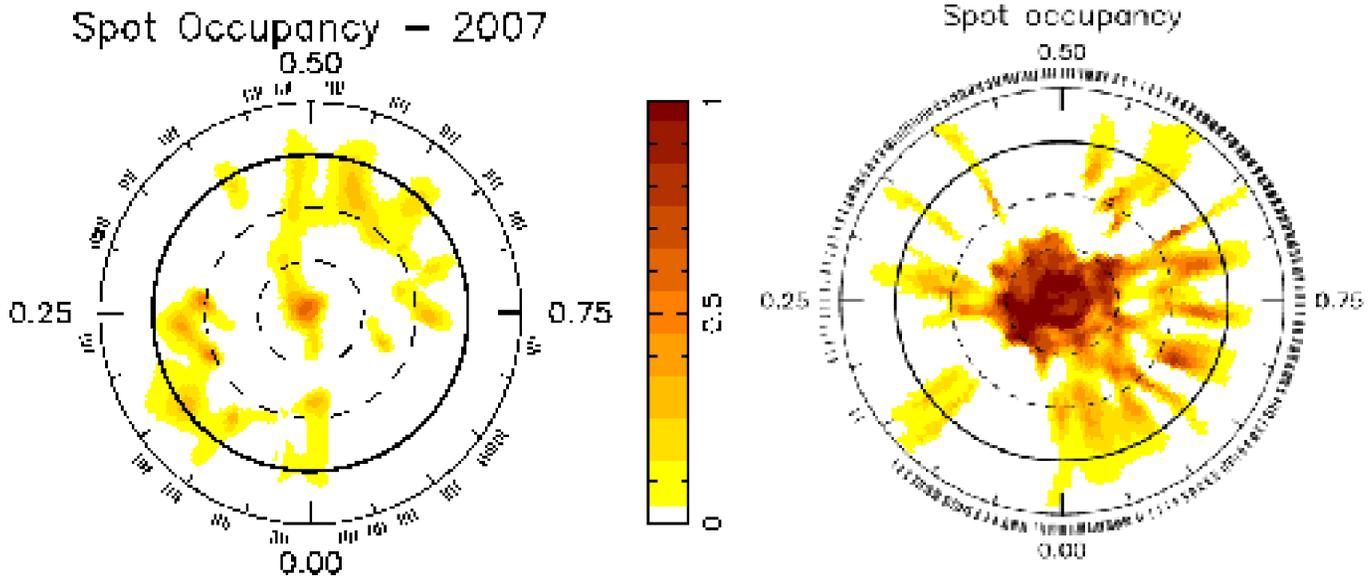}
\caption{Doppler image of the G dwarf HD 141943 (left) and the K dwarf AB Dor (right)}  
\protect\label{f-DopplerImages} 
\end{center}
\end{figure*}

Observations of rapidly rotating young G and K dwarfs consistently show starspot distributions that predominantly comprise high latitude or polar spots.  For example, a typical starspot pattern that has been derived for the G2 dwarf HD 141943 \citep{Marsden2011} and a surface image for the K1 dwarf AB Dor \citep{Jeffers2007} are shown in Figure~\ref{f-DopplerImages}.  

\begin{table}
\begin{tabular}{l c c c c}
\hline 
Stellar Parameters: & AB Dor (Ref) & HD141943 (Ref) \\ 
\hline
Spectral Type & K0 V  & G2 V \\
T$_{phot}$ & 5000 K (1) & 5850 K (2) \\
T$_{spot}$ & 3500 K (1) & 3950 K (2) \\
Period & 0.51479-d (2) & 2.182-d (4)\\
Inclination & 60$^\circ$ $\pm$ 5$^\circ$ (3) & 70$^\circ$ $\pm$ 10$^\circ$ (4)\\
$v$ sini $i$ & 89 $\pm$ 0.5km s$^{-1}$ (3) & 35.0 $\pm$ 0.5 km s$^{-1}$ (4)\\
\hline
\end{tabular}
\caption{Stellar parameters for AB Dor and HD141943. References: $1$ Donati et al. (2000); $2$ Innes et al. (1988); $3$ Jeffers et al. (2007); $4$ Marsden et al. (2011)}
\protect\label{t-stellarparameters}
\end{table}

In contrast to the Sun, where starspots are located at equatorial latitudes, young solar-type stars typically show starspots at high latitudes including polar spots.  This has been consistently imaged for more $\sim$ 75 late-type stars by different research groups using different tomographic imaging techniques \citep[e.g. summary by][]{Strassmeier2009}.  The fundamentally different distributions of starspots on the Sun and young late-type stars has been interpreted as being evidence for a different dynamo regeneration process occurring in much younger solar-type stars.  This is supported by observations of the magnetic field topology on young solar-like stars \citep{Jeffers2011,Marsden2011} where they consistently show large rings of azimuthal field located around the star's rotation axis.  Such magnetic field distributions are also in contrast to the magnetic field topology observed on the Sun, where additionally, the most prominent evidence of the magnetic field regeneration is the reversal of the magnetic field every 11 years.  However, observations of young stars over a long-time span have yet to show evidence for such cyclic magnetic activity. \newline

In the first paper of this series \citep{Barnes2011} we showed that the distortion caused by a starspot in the star's radial velocity profile is dependent on the latitude of the starspot, with spots at equatorial latitudes giving a much stronger distortion compared the same spot at high latitudes.  If indeed all young active solar-type stars have strong polar caps then the distortion caused by the spots on the star's radial velocity profile will be different than if the spots are located at equatorial latitudes as observed on the Sun or indeed homogeneously distributed over the star's surface.

In this paper we will use the reconstructed images of AB Dor and HD 141943 as representative distributions of starspots for K dwarfs and G dwarfs.  The stellar parameters of these two sources are summarised in Table~\ref{t-stellarparameters}.  While other tomographic images of similar stars show the same global features, any smaller scale deviations from the adopted spot distributions are considered as uncertainties in the model setup.





\subsection[]{Plage on G and K dwarfs}


Additional distortions on a planet's radial absorption line profile can be caused by bright stellar plage.  The distributions of plage on active stars, are unfortunately not well understood, though several constraints can be placed from long-term monitoring of active young stars.  An example is the work of \cite{Radick1998} who contemporaneously monitored the chromospheric Ca H\&K emission and photometric brightness levels of a sample of 35 stars.  \cite{Radick1998} and later \cite{Lockwood2007} conclude that the younger and more active stars of their sample become fainter as their Ca HK emission increases, in contrast to what is observed on the Sun \citep{Unruh1999}.  This would imply that older stars such as the Sun become plage dominated at activity maximum, while younger stars such as HD 141943 and AB Dor become spot dominated at activity maximum.

To determine realistic distributions of plage on the surfaces of our modelled young active G and K-dwarf stars we use the results of \cite{Unruh1999} as input to our model.   In this work we assume that a plage region is located around a starspot with a radius of twice the spot radius. 

\subsection[]{Random spots}

Doppler imaging studies have revealed spot coverage fractions of the order of $\approx$ 10\%, while other methods such as TiO band analysis suggest that the starspot coverage fraction could be as high as 50\% in active G and K dwarfs \citep{oneal1998, oneal2004}.  The discrepancy between the spot coverage fractions derived using these two independent methods can be resolved if the stars exhibit many small peppered spots that are too small to be resolved using the Doppler imaging technique.  

To account for this we model additional starspot coverage in addition to the large-scale starspots reconstructed using the Doppler imaging tomographic imaging technique.  The distributions of small peppered starspots are modelled to follow the log-normal size distribution size of dark spots on the Sun \citep{bogdan1988} and extrapolated to active stars by \cite{Solanki1999} and modelled following the same methods as \cite{Jeffers2005} and \cite{Barnes2011}.  The peppered spots are applied to the surface of an immaculate star and are distributed in (i) longitude, randomly between 0$^\circ$ and 360$^\circ$ (ii) latitude, following ($\theta = arcsin(2x+1)$) with 0$\leq$x$\leq$1 to eliminate an artificial concentration of spots at the pole (iii) spot radius, computed using the log-normal distribution of \cite{bogdan1988} and (iv) spot brightness \& spot sharpness, which are modelled to obtain a umbral to penumbral ratio of 1:3 \citep{Solanki1999}.  Further details are described in \cite{Jeffers2005} and \cite{Barnes2011}.  

\subsection{Simulated activity models}
\protect\label{s-models}

Using the individual tomographic images as a starting point, the final spot model and plage distributions are modelled on an immaculate star as representative activity distributions for young G dwarfs and K dwarfs in V, I and YJH wavelength bands.  To understand the impact particularly of each activity signature on the radial velocity profile, we model the following nine cases.  The wavelength bands are chosen to be comparable to those used in upcoming planet searching surveys such as CARMENES \citep{Quirrenbach2012}.

\begin{itemize}
\item [\bf 1.] {{\bf Spots only}} -- modelled from the tomographic Doppler images of HD 141943 and AB Dor - V-band
\item [\bf 2.] {{\bf Spots + Plage }} -- as for case 1 but with the addition of large regions of plage surrounding starspots - V-band
\item[\bf 3.] {{\bf Spots + Random Spots}} -- case 1 with many small random spots - V-band
\item[\bf 4.] {{\bf Spots + Plage + Random Spots}} -- case 2 with many small random spots - V-band
\item[\bf 5.] {{\bf I-band Spots + Random Spots}} -- case 3 but for I-band
\item[\bf 6.] {{\bf I-band Spots + Plage + Random Spots}} -- case 4 but for I-band
\item[\bf 7.] {{\bf YJH-band Spots + Random Spots}} -- case 3 but for YJH-band
\item[\bf 8.] {{\bf YJH-band Spots + Plage + Random Spots}} -- case 4 but for YJH-band
\item[\bf 9.] {{\bf Immaculate Star:}} as above, but without any stellar activity - V-band
\end{itemize}

The total modelled spot filling factors are for the K dwarfs: 10.3\% (observed Doppler image) and 31\% (observed Doppler image + Random spots) and for the G dwarfs: 7\% (observed Doppler image) and 30\% (observed Doppler image + Random spots).  
Plage, which is added with radii = 2 * spot radius (from the Doppler image) is 11\% for the K dwarfs and 6\% for the G dwarfs.


\begin{figure*}
\begin{center}
\includegraphics[angle=270,scale=0.7]{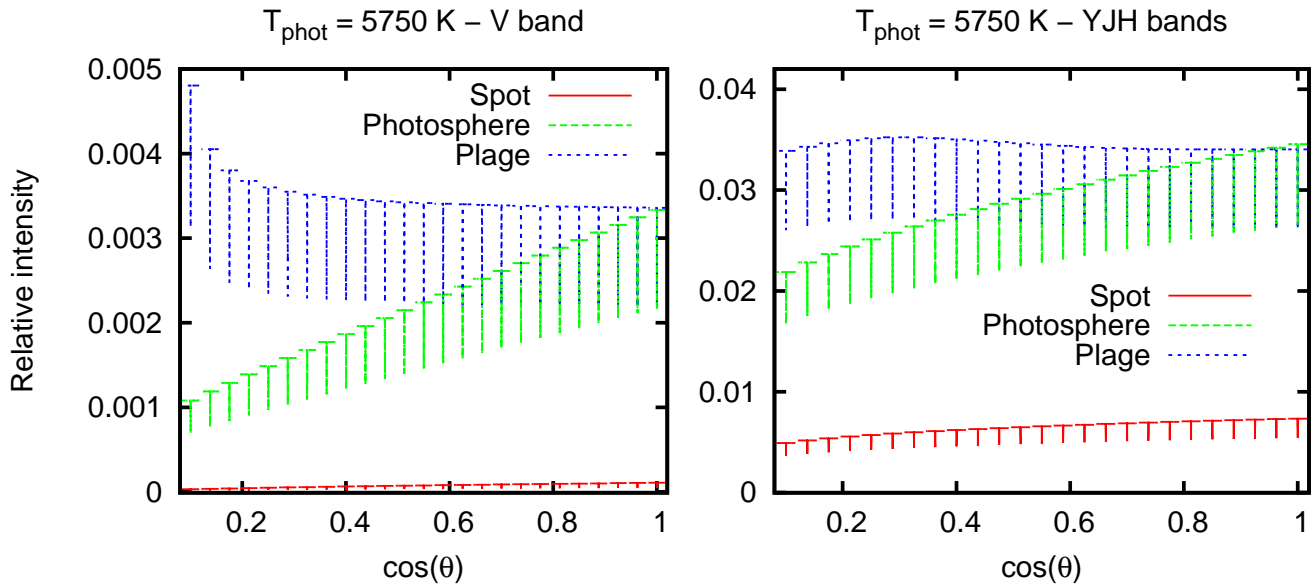}
\caption{The centre to limb variations used in this analysis for Gdwarfs for V-band (left) and YJH bands (right) with photospheric temperature = 5750 K and spot temperature = 4000 K. The plage I-band centre to limb variation follows the same limb brightening as as the V-band wheres the infrared YJH variation reaches a peak at cos($\theta$) $\sim 0.3$. The line depths are derived from {\sc vald} line lists, and represent the mean line depth over each wavelength regime. In our models, it is the relative intensities that are important. The plots illustrate the decrease in contrast between spots and photosphere in the infrared YJH band as compared with the V band. In this instance the continuum ratios at disk centre, I$_{0,\rm phot}$/I$_{0,\rm spot}$ are 30.9 and 4.7 for V and YJH respectively.}
\protect\label{f-centre_to_limb}
\end{center}
\end{figure*}

\section{Stellar line profiles}

Stellar line profiles are generated for each of these activity models using the imaging code DoTS ``Doppler Tomography of Stars" \cite{Colliercameron1997}.  DoTS uses a 3D stellar model that has been recently updated to model both cool dark stars spots and regions of bright plage.  The relative fluxes for a star and its starspots and hotspots are determined from the absolute and magnitude and radii models of \cite{Baraffe1998}. 

\subsection{Centre to limb intensity variations}
\protect\label{section:CTLV}

Non-linear limb-darkening values from \cite{claret2000} are used to define the radial variation in intensity from the centre of the star to its limb for both photospheric and spot intensity levels.  An important consideration of including plage in our model is that solar plage is shown to exhibit limb brightening, and not limb darkening as in the case of the stellar photosphere and starspots.   The variation of plage temperature as a function of limb angle is taken from Figure 3 of \cite{Unruh1999} for a wavelength of 524.5 nm using their limb brightening law of delta I / I = b (mu$^{-1}$ -a).  A similar approach has been used by \cite{Meunier2012}. Figure \ref{f-centre_to_limb} (left panel) shows the centre to limb variation of our three-temperature model, illustrating the limb darkening in the spots and photosphere and the limb brightening in the plage. The line depth used in each case is derived from the Vienna Atomic Line Database {\sc vald} \citep{Piskunov1995, Ryabchikova1997,Kupka1999,Kupka2000} for the wavelength regions considered. With no studies at red-optical wavelengths, we have adopted the same limb brightening law as those centered around V band. At 1.6 $\mu$m, the centre to limb variation variation is different from that found at optical wavelengths \citep{SanchezCuberes2002,Fontenla2004}. We have modelled our infrared centre to limb variation, shown in  Figure \ref{f-centre_to_limb} (right panel), on the empirical results of \cite{SanchezCuberes2002}.


\subsection{Temperature Planes}

Three temperature planes for the stellar photosphere, hotspots and coolspots are used and are shown in Figure ~\ref{f-centre_to_limb}.  An important consideration in the modelling of starspots is the photospheric/spot temperature contrast ratio.  To understand the impact of this parameter we have taken the two extremes found in the literature: T$_{spot1}$ = 0.65 T$_{phot}$ and T$_{spot2}$ = T$_{phot}$-1500 K.  This results in spot temperatures, for the G dwarf, of T$_{spot1}$ = 0.65 T$_{phot}$=4000 and T$_{spot2}$=4250 K, while for the K dwarf, using the same equations T$_{spot1}$=3250 K for ratio 1 and 3500 K using ratio 2.  The temporal variation of plage throughout the stellar magnetic cycle is beyond the scope of this work and will be addressed in a future paper.




\begin{figure}
\begin{center}
\includegraphics[angle=0,scale=0.49]{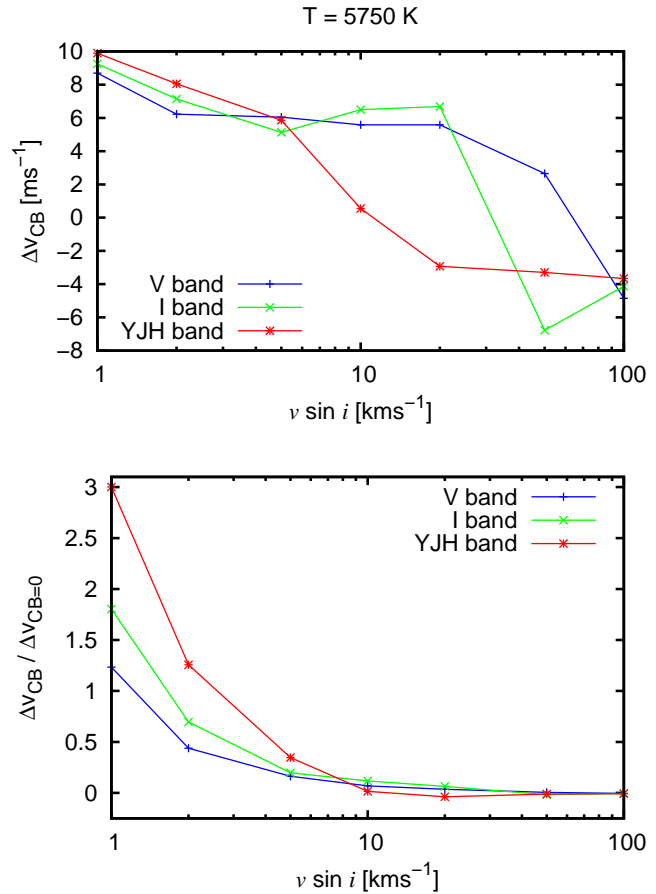}
\end{center}
\caption{Top: The convective blueshift (CB) component of the radial velocity for our T = 5750 K, G dwarf with DI + plage. The plot shows the {\em half amplitude} difference for the cases with a CB equal to 525\,-\,600 ms$^{-1}$, as indicated in the main text, and the equivalent cases with no blueshift. Bottom: The fractional modification of the radial velocity amplitude with CB compared with the case with no CB.}
\protect\label{fig:convblue}
\end{figure}

\subsection{Convective blueshift}
\protect\label{section:blueshift}

Recent modelling of solar radial velocity jitter has incorporated the effects of convective blueshift \citep{Meunier2010}. The convective blueshift (CB) effect is found to induce variations of up to $\sim 10$ ms$^{-1}$ during a complete solar cycle, as a result of magnetic activity phenomena such as spots and plage. Hence for completeness, we include and asses this effect in our model. The CB in an immaculate star is a well established limiting factor in radial velocity precision \citep{Dravins1982}. An overall distortion of the stellar line occurs as a result of the local intensity profile from each convective granule comprising different components. Since the centre of each convective cell, is brighter than the outer regions of the cell and the intergranular lanes, these regions contribute a greater intensity to the disk integrated line profile. Any difference in the velocities between the convective cell centre and intergranular lanes will thus lead to a local intensity profile with distorted line symmetry, leading to an effective blueshift of the line. The disk integrated line profile therefore also possesses this apparent asymmetry. The effect is greatest at disk centre where the full component of the radial convective motions is seen, and falls to zero at the limb. In the starspots, the relative intensity is much weaker as illustrated in Figure \ref{f-centre_to_limb}). The magnetically isolated spots suffer much less from the CB effect as a result of the inhibited convective motion. Hence while the starspots themselves distort the line profile because of the lower intensity within the spots, the local intensity profile of a spot also appears slightly redshifted because of the lack of CB that is seen in quiet photosphere. Plage also masks the CB of the photospheric regions, attenuating it by a factor of 1/3 \cite{Brandt1990}. The magnitude of the CB is of order 500 ms$^{-1}$ \cite{Gray2009} but ranges from 200 ms$^{-1}$ in K dwarfs to 1000 ms$^{-1}$ in F dwarfs \citep{Dravins1990}. 


The asymmetry of the photospheric line profiles is not expected to induce RV variations on either a quiet star, or equivalently, a model with no magnetic features (either cool spots, or plage regions). We have therefore incorporated the CB effect by modelling a shift in the starspot profile (which is positive, i.e. a redshift) relative to the photospheric profile. For umbral regions, we have estimated the CB for our simulated G2V star from Gray (2009) who measured the line shift as a function of line depth. Since smaller lines form at deeper levels, they are subject to higher CBs that stronger lines. Using mean line depths derived from {\sc vald} (\S \ref{section:CTLV}), we estimate the relative shift of starspots for both our K and G stars, and incorporate these values into our model. Since we use a three temperature model and filling factors (ranging from zero to unity), the model simulates a linear interpolation of the spot shift. Hence a model pixel with complete spot filling would see the full CB removed (or equivalently, the maximum spot redshift), while a filling factor of 0.5 would only experience half the maximum shift. For the G dwarf (T = 5750 K) model, assuming that the blueshifts in Gray (2009) can be extrapolated to the red-optical and infrared, and that only line depth changes are important, we estimate CBs of -525, -600 \& -600 ms$^{-1}$ for V, I and YJH bands respectively. Similarly for our K dwarf (T = 5000 K) model, we estimate CBs of -200, -225 and -225 ms$^{-1}$.

To investigate the effect of CB in our simulations, we computed the line profiles and cross-correlation velocities for the simulated range of $v$\,sin\,$i$ values both with, and without the the blueshift effect. Figure \ref{fig:convblue} (upper panel) shows the full amplitude of the velocity variations induced in the V band by our T = 5750 K star with spots from Doppler imaging and simulated plage (Model: DI+pl). In the V band, we find that rotation induces CB velocities ranging from 8.7 ms$^{-1}$ at $v$\,sin\,$i$ = \hbox{1 kms$^{-1}$} to \hbox{-4.8 ms$^{-1}$} at $v$\,sin\,$i$ = \hbox{100 kms$^{-1}$}. Since the amplitude of the blueshift relative to the $v$\,sin\,$i$ is greatest for the lower $v$\,sin\,$i$ values, the resulting jitter is most pronounced. With increasing $v$\,sin\,$i$, the effect becomes gradually smaller. In other words, as the spot-induced radial velocity jitter increases, the relative, effect of the CB becomes smaller. However, as $v$\,sin\,$i$ increases, more spot features also become resolved with the line profile. The relationship between increasing $v$\,sin\,$i$ and $\Delta v$ shown in Figure \ref{fig:convblue} (upper panel) is thus not linear, and in the illustrated case has led to a reduced jitter for $v$\,sin\,$i$ = 100 kms$^{-1}$. In Figure \ref{fig:convblue} (bottom), we show the fractional change in the velocity amplitude due to spots as a result of including CB, emphasising the decreasing importance with increasing $v$\,sin\,$i$, varying from 123 per cent to -0.5 per cent. The effect is greater at longer wavelengths, because the relative radial velocity augmentation due to the CB is larger relative to the radial velocity jitter of the spots, which become weaker with increasing wavelength.

Figure \ref{fig:convblue} also illustrates that the wavelength dependence of observations does not follow a simple augmentation of the CB effect as a function of $v$\,sin\,$i$. This is likely due to a combination of the above mentioned increasing resolution, but also the change in relative strength of the spot and plage contributions (in the case of the plage, the centre to limb variation is also different as discussed in \S \ref{section:CTLV}). While the photosphere/spot contrast becomes smaller with increasing wavelength, the plage is simulated with the same relative central brightness in V and I bands and thus the effects of plage become more dominant than at shorter wavelengths. In reality, the behaviour of the CB effect as a function of spot and plage morphology and wavelength is not straightforward and each case must be taken on an individual basis.

Despite the active nature of our simulated stars, we find a comparable V band CB amplitude to that found for the Sun by \cite{Meunier2010}. This is likely due to the high latitude of the larger spots and plage regions used in our Doppler images since they do not pass through the disk centre where the CB is maximised.  


\subsection{Stellar \vsinis}

In their early evolution, solar-type stars (M\subs{\star} $\ga$ 0.5 \Msolar) undergo contraction that acts to spin-up the rotation rate of the star \citep{Bouvier2007}. Consequently, young solar-type stars can evince a large range of rotation rates, with projected rotational velocities (\vsinis values) over 200 \kmss \citep{Marsden2009}. The peak rotation rate a young star reaches is heavily dependent on the disc-locking lifetime of the young star \citep{Bouvier2009}.  For most solar-type, zero-age main-sequence (ZAMS) stars such as HD 141943 and AB Dor, the peak \vsinis reached is $\la$ 100 \kmss (although a few have rotation rates over 200 \kmsn) as shown by the \vsinis values of the solar-type stars in the ZAMS-age young open clusters IC 2391 \& IC 2602 \citep{Marsden2009}. These results show that almost all of the G-stars in the two clusters have a \vsinis $\la$ 100 \kmss with a large scatter in the \vsinis values of the stars. For the K-stars in the clusters, the majority are also rotating at or below 100 \kmsn, with only a very few rotating more rapidly than this.  As the stars then evolve on the main-sequence their rotation rates gradually drop due to magnetic braking \citep{Bouvier2009} until, by the age of the Hyades ($\sim$625 $\pm$ 50 Myrs, \citealt{Perryman1998}), the rotation rates of solar-type stars are very much linked to their spectral-type with higher-mass stars having shorter periods than lower-mass stars \citep{Gray2008}.


In this work we include a large range of \vsinis values, with constant activity levels, to quantify how \vsinis impacts planet detection thresholds for zero age main sequence stars with high activity levels.  Our aim is to understand whether stellar \vsinis or activity levels are the main limitation of detecting planets around young active stars.

\section{Modelling Radial Velocity Data}

\begin{figure*}
\begin{center}
\subfigure[K dwarfs rms scatter]{\includegraphics[angle=270,scale=0.34]{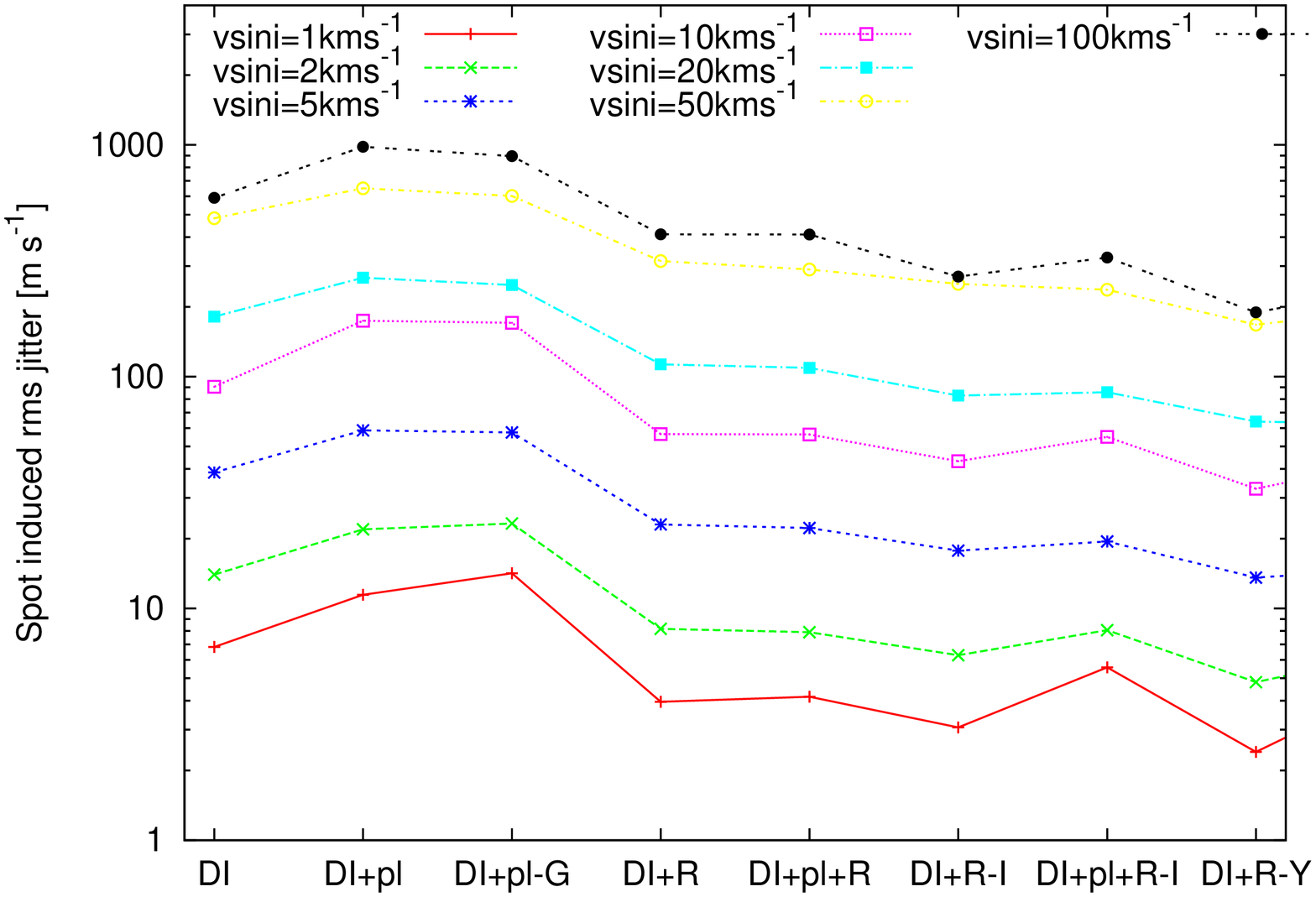}}
\subfigure[G dwarfs rms scatter]{\includegraphics[angle=270,scale=0.34]{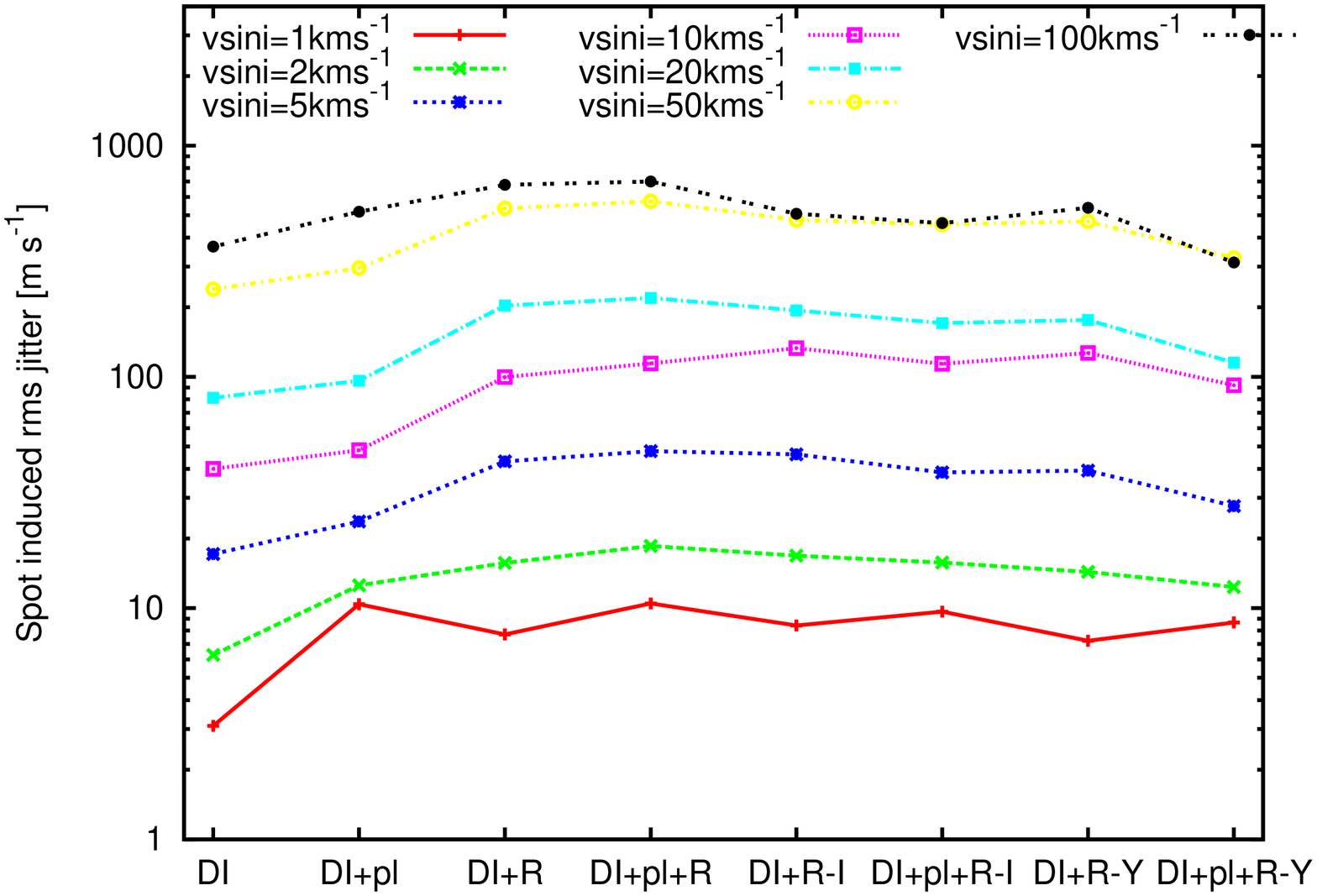}}
\caption{The RMS radial velocity variation for K and G dwarfs (right and left panels respectively) each of the 7 models:  DI: Doppler image; DI+pl: Doppler image + Plage; DI+pl-G: K dwarf spots on Gdwarf; DI+R: Doppler image + Random Spots; DI+pl+R: Doppler image + Plage + Random Spots; DI+R-I: I-band Doppler image + Random Spots ; DI+pl+R-I: I-band Doppler image + plage + Random Spots; DI+R-Y: YJH-band Doppler image + Random Spots ; DI+pl+R-Y: YJH-band Doppler image + plage + Random Spots.  The radial velocity variation for Immaculate star = 0 ms$^{-1}$}.  
\protect\label{f-rms-maxjitter} 
\end{center}
\end{figure*}

\begin{figure}
\begin{center}
\includegraphics[angle=270,scale=0.32]{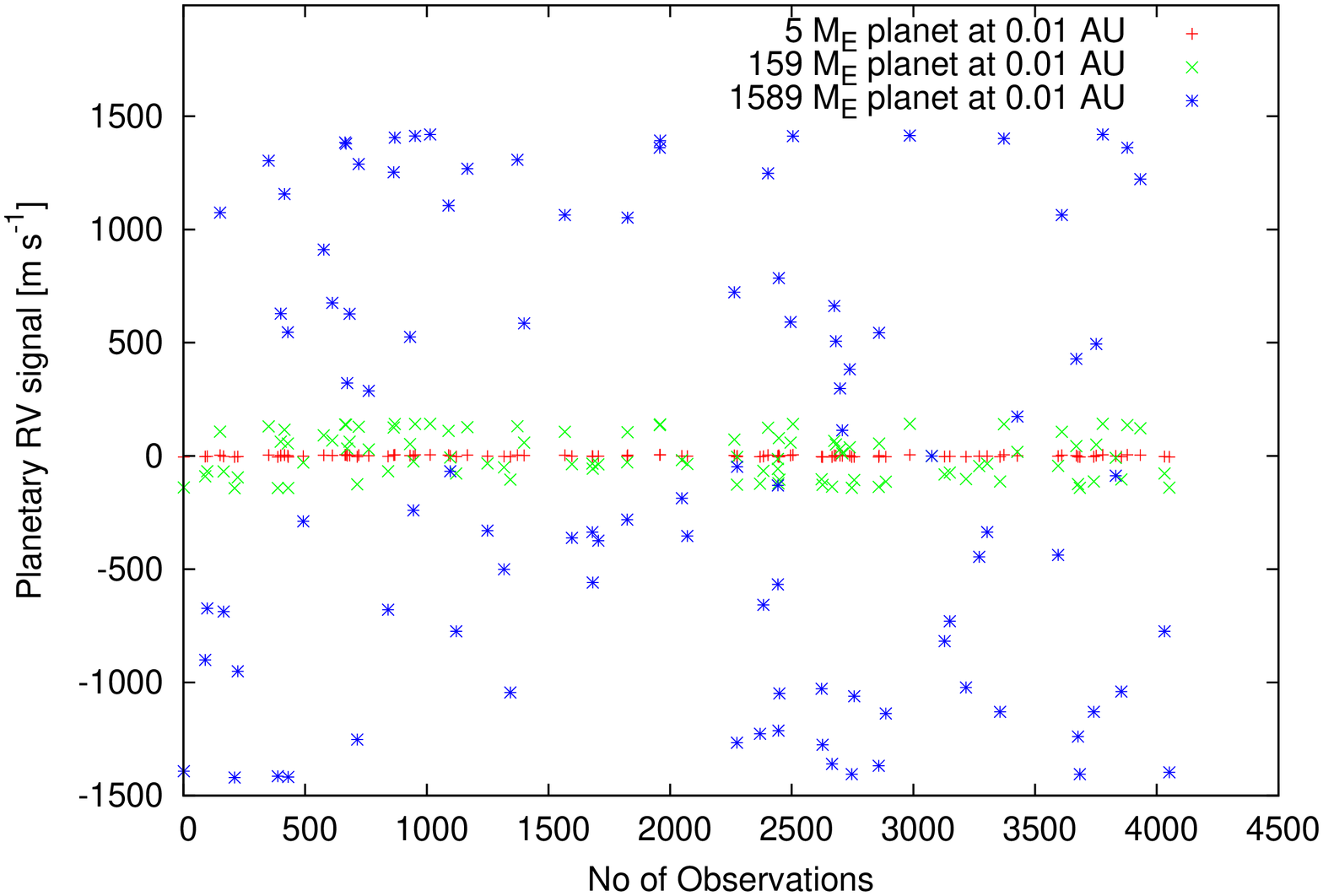}
\includegraphics[angle=270,scale=0.32]{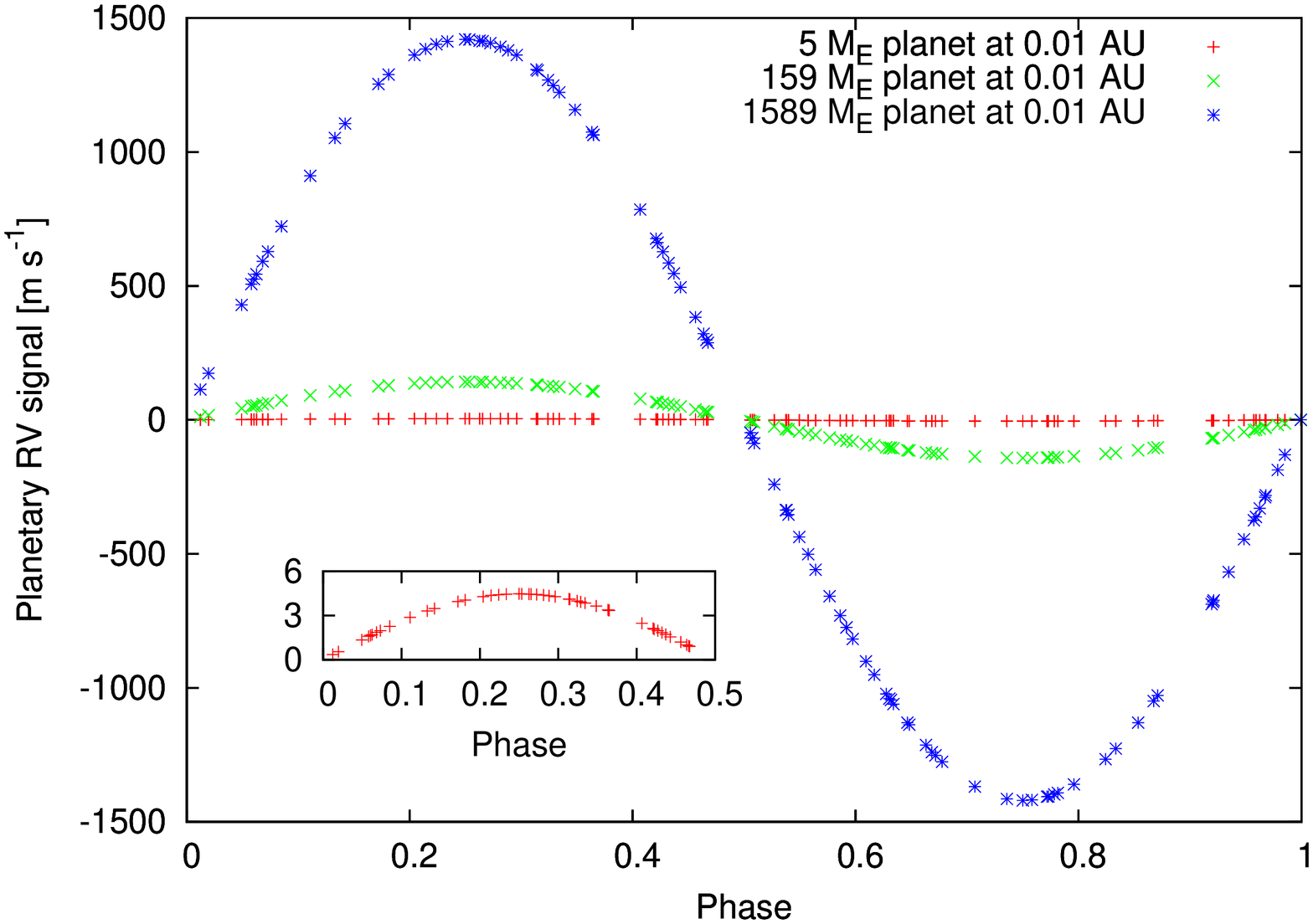}
\caption{The planet induced radial velocity.  Upper panel: unphased data for planetary masses of 5 M$_{\oplus}$, 159 M$_{\oplus}$ (0.5 M$_{\mathrm{J}}$) and 1589 M$_{\oplus}$ (5 M$_{\mathrm{J}}$).  Lower panel: phased data from upper panel with the inset showing the radial velocity variation for M$_{planet}$ = 5 M$_{\oplus}$.}
\protect\label{f-planetaryphase}
\end{center}
\end{figure}

In this Section we simulate a stellar radial velocity profile, for a G and a K dwarf star, to investigate the radial velocity amplitudes resulting from non-uniform line profiles.  The stellar radial velocity profile comprises (1) the jitter caused by activity (starspots and plage) on the host star (2) the radial velocity induced by a planet orbiting its host star and a contribution from (3) the achievable instrumental precision. 

\subsection{Radial velocities induced by activity jitter}

Radial velocity measurements are made directly from the line measurements as described by \cite{Barnes2011} and are briefly summarised here.  Stars with an axial inclination, $i$=90$^\circ$ are simulated according to the parameters in Table~\ref{t-stellarparameters} for each of the 9 stellar models described in Section~\ref{s-models}.  A total of 90 line profiles (i.e. every 4$^\circ$) with a range of $v$sin$i$ values of 1, 2, 5, 10, 20, 50 and 100 km s$^{-1}$ are generated for a complete stellar rotation period.  A synthetic Voigt profile is used to simulate the line profile of a non-rotating star, with a resolution of 70,000 to be consistent with the simulations in paper I \citep{Barnes2011} and comparable to the spectral resolution of CARMENES \cite{Quirrenbach2012}.


The variation of spot induced jitter per model is plotted in Figure~\ref{f-rms-maxjitter} for K and G dwarfs.  For the K dwarf (left panel), there is an increase in the activity induced jitter when plage is added to the model with only the Doppler image (Model DI+pl).  However when random spots are added to the model (Model DI+R) there is a surprising decrease in the activity induced jitter.  This is due to the placing of random spots at longitudes that exhibited immaculate photosphere in the Doppler images.  The result is that the phased induced asymmetries, while more numerous, are more balanced, leading to reduced radial velocity information.  The addition of plage to this model has little impact (Model DI+pl+R).  For the G dwarf (right panel) the plot clearly shows that the most heavily spotted stars have the most starspot induced radial velocity jitter at all wavelength bands. The addition of plage has a less dramatic increase for the G dwarf as it only covers 6\% of the stellar surface in contrast to 11\% for the K dwarf.  Since there was no activity induced jitter for the case of the immaculate star, this is not shown.  In general, the levels of activity induced rms jitter are comparable with the work of \cite{Barnes2011} for their very active stars, though they only considered a maximum $v$sin$i$ value of 50 km s$^{-1}$.  As we found in \cite{Barnes2011}, the latitude of the starspot is related to the amplitude of the distortion in the radial velocity curve, with spots at lower latitudes having a stronger impact than spots at high latitudes.  The models with random spots contain many more spots at lower latitudes therefore producing a stronger distortion, which can also have the effect of negating other activity distortions.  As we have used two different input Doppler images it is not possible to directly compare the activity induced jitter for the K and G dwarfs.

\subsubsection{Correlation of activity jitter with \vsinis}

Additionally from the two panels of Figure~\ref{f-rms-maxjitter} there is a direct correlation between spot and plage induced activity jitter and the stellar \vsinis for both the K and the G dwarfs.  A doubling of the stellar $v$sin$i$ clearly has the impact of doubling the activity induced rms jitter.  This relation is valid for values of $v$sini$i$ of up to 50 km s$^{-1}$ where it begins to flatten out due to the activity features being more clearly resolved at extreme $v$sin$i$ values.  Indeed the inclusion of $v$sin$i$ = 100 km s$^{-1}$ is quite unrealistic for planet hunting, though it is simulated for illustrative purposes. The correlation between activity induced jitter and stellar $v$sin$i$ is in agreement with previous results for M dwarfs \citep{Barnes2011}.  In Figure~\ref{f-rms-maxjitter} it is clearly evident that the precise origin of stellar activity (i.e. plage, spots, etc) can be an order of magnitude less important than the stellar \vsinis.  In contrast to the results of \cite{Barnes2011}, we find that the spot/photospheric temperature contrast ratio used has an insignificant impact on the activity induced jitter values.



\subsection{Radial velocities induced by planets}

The planetary mass, M$_{planet}$ is set to be equal to 1, 2, 5, 10, 20, 50 and 100
M$_{\oplus}$ and 1, 2 and 5 M$_{\mathrm{J}}$ orbiting with an orbital radius, a$_{planet}$ = 0.01, 0.02, 0.05, 0.1, 0.2, 0.5, 1, 2 and 5 AU. These parameters were chosen as a representative sample of planetary
mass and orbital radii from the catalogue of detected exoplanets [www.exoplanet.eu].  Table~\ref{t-KG_gdwarf} summarises these parameters and also lists the corresponding stellar radial velocity amplitudes K$_\star$ for each planet.  All planetary orbits are modelled with an inclination angle = 90$^\circ$.  An example of the planet induced radial velocity is shown in Figure~\ref{f-planetaryphase}.

\begin{figure*}
\begin{center}
\subfigure[Doppler image]{\includegraphics[angle=270,scale=0.22]{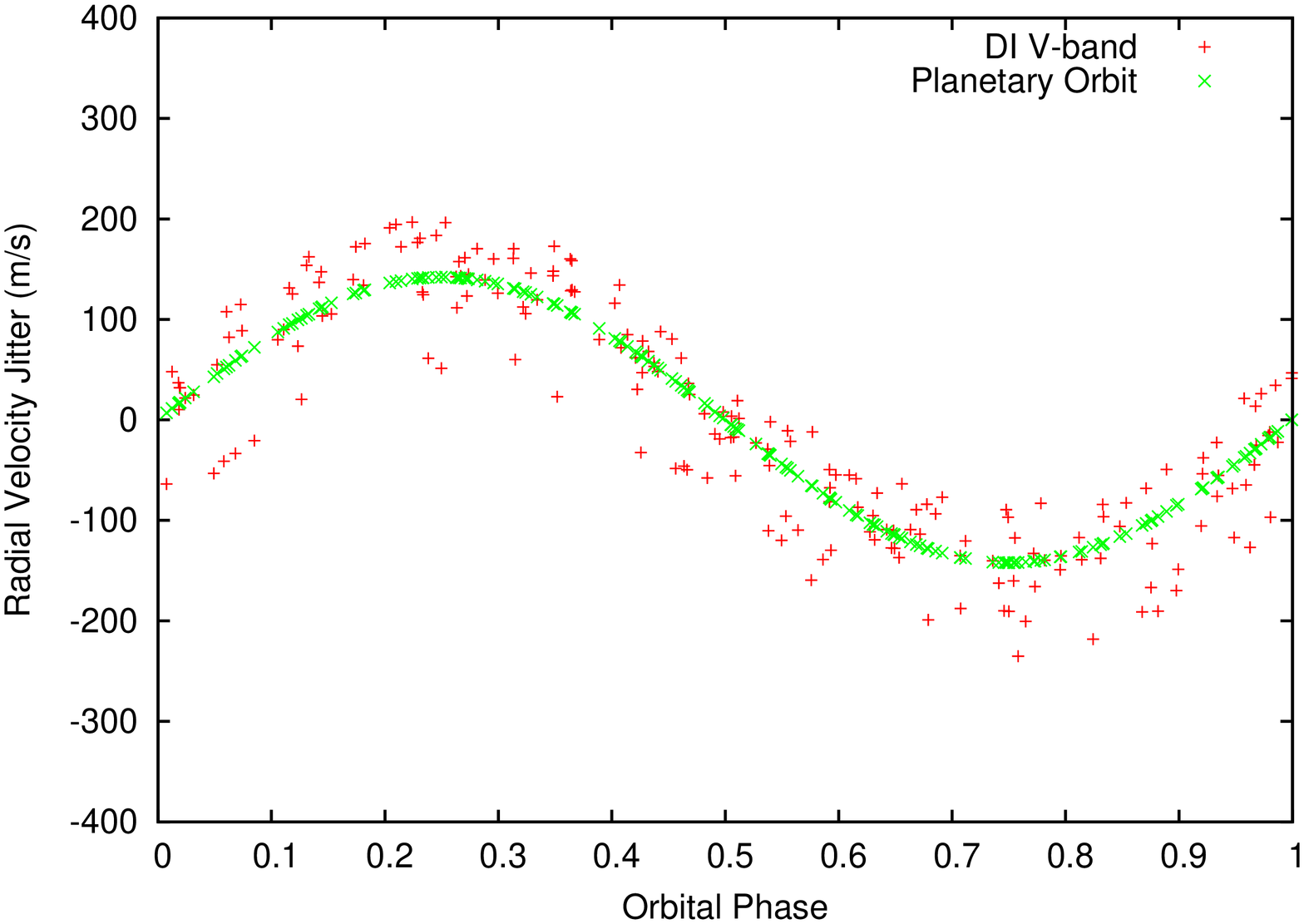}}
\subfigure[Doppler image + Plage]{\includegraphics[angle=270,scale=0.22]{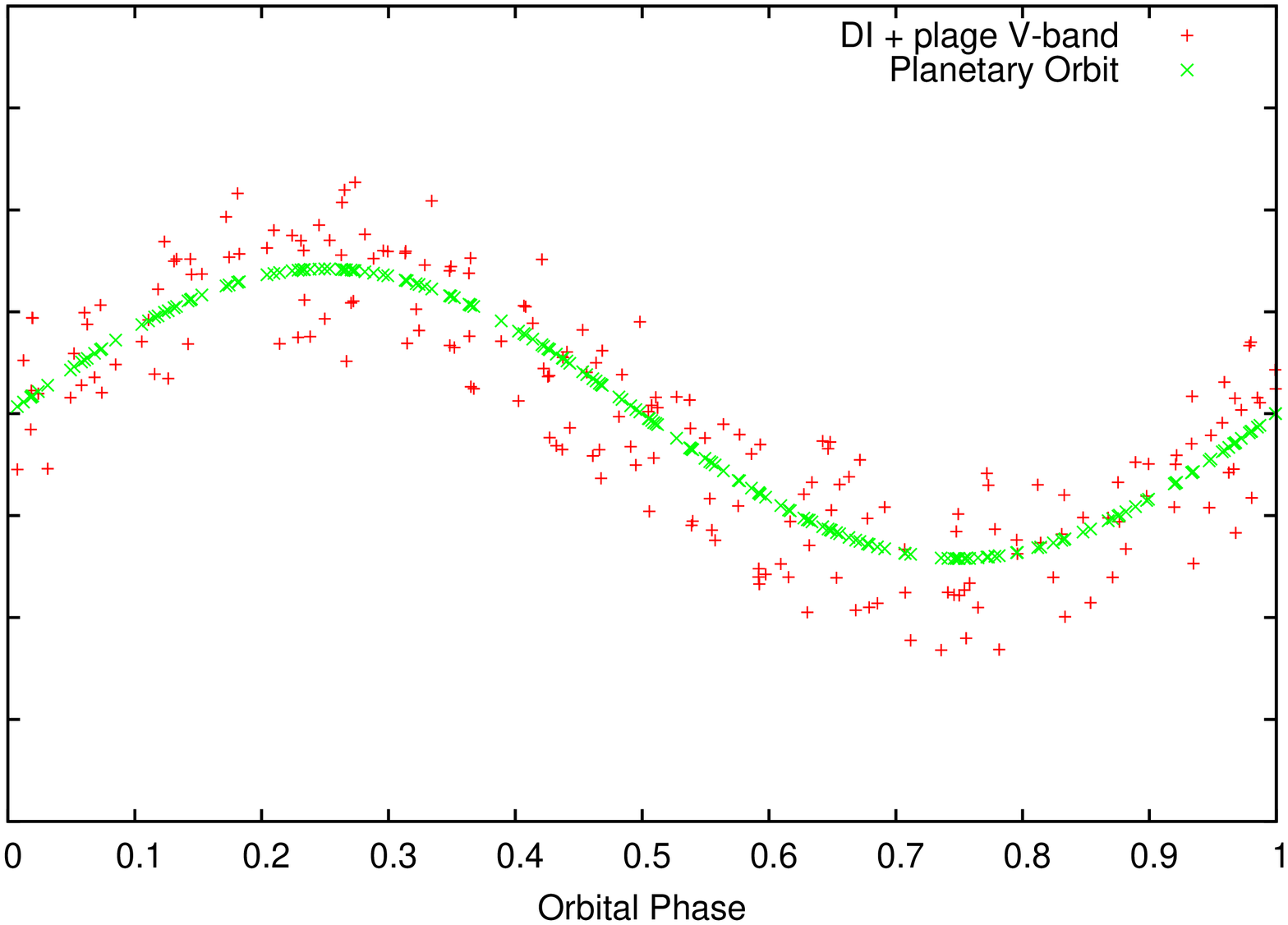}} 
\subfigure[case (a) + Random Spots]{\includegraphics[angle=270,scale=0.22]{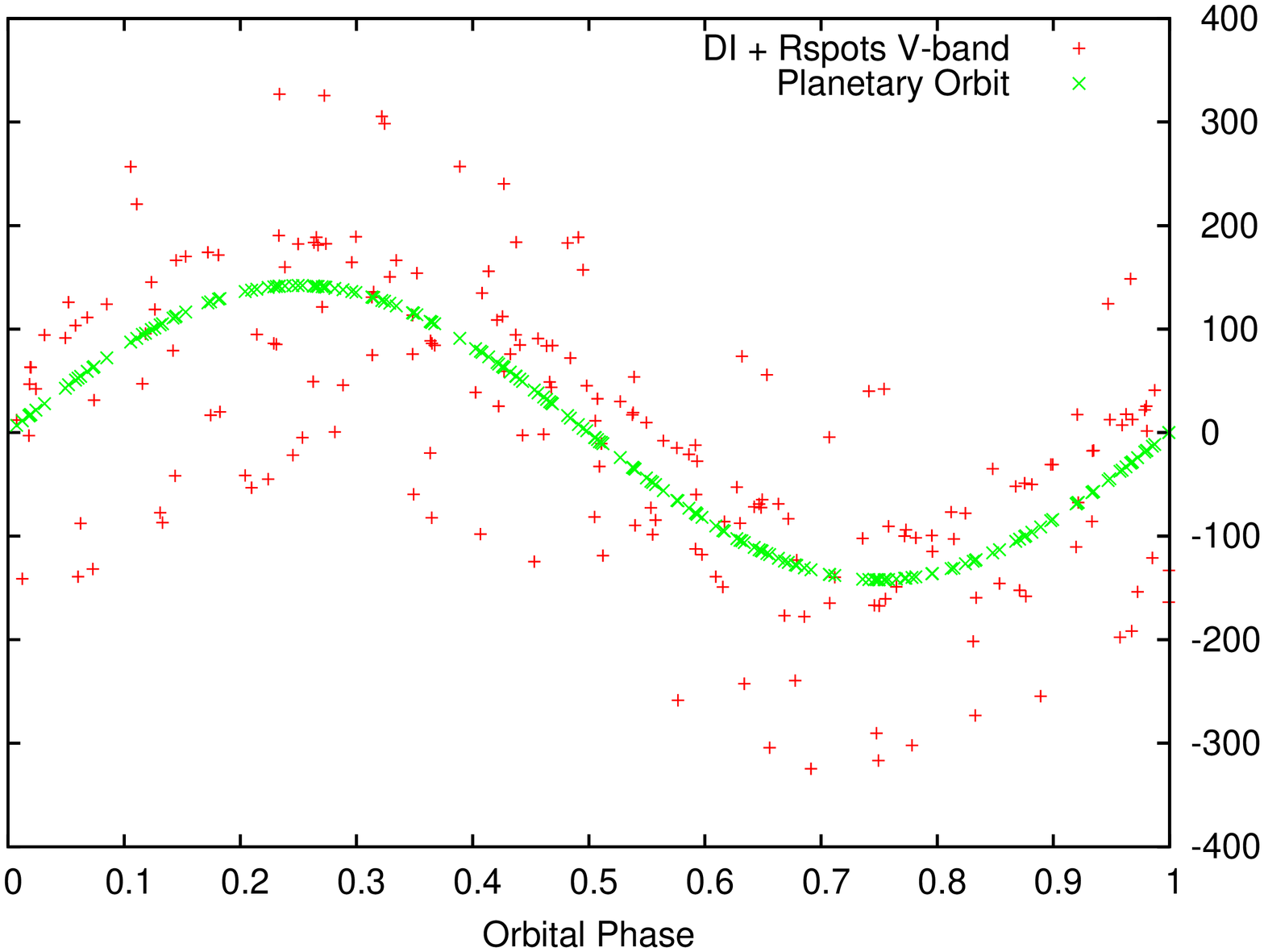}}
\subfigure[case (b) + Random Spots]{\includegraphics[angle=270,scale=0.22]{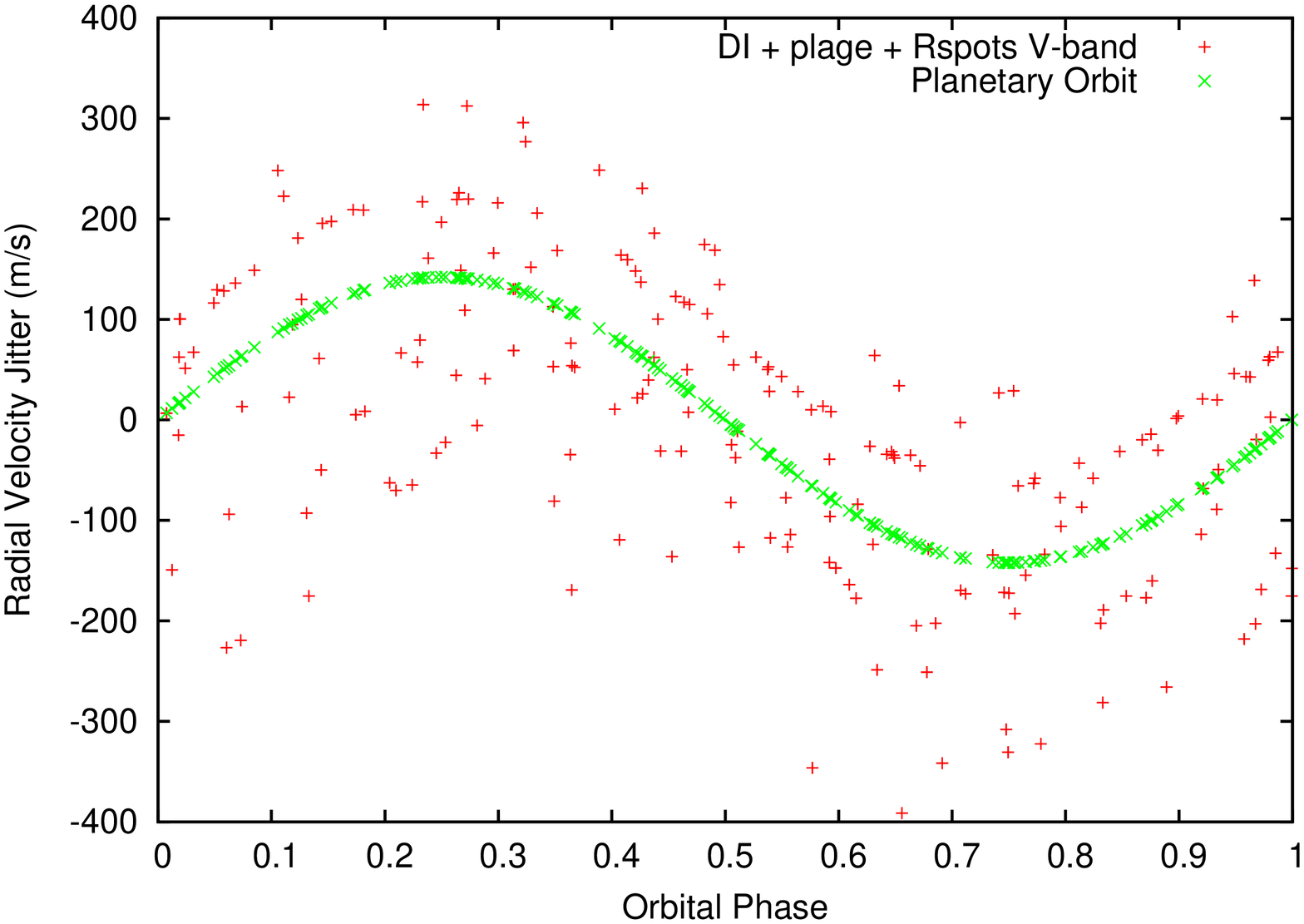}}
\subfigure[Doppler image + Random Spots I-band]{\includegraphics[angle=270,scale=0.22]{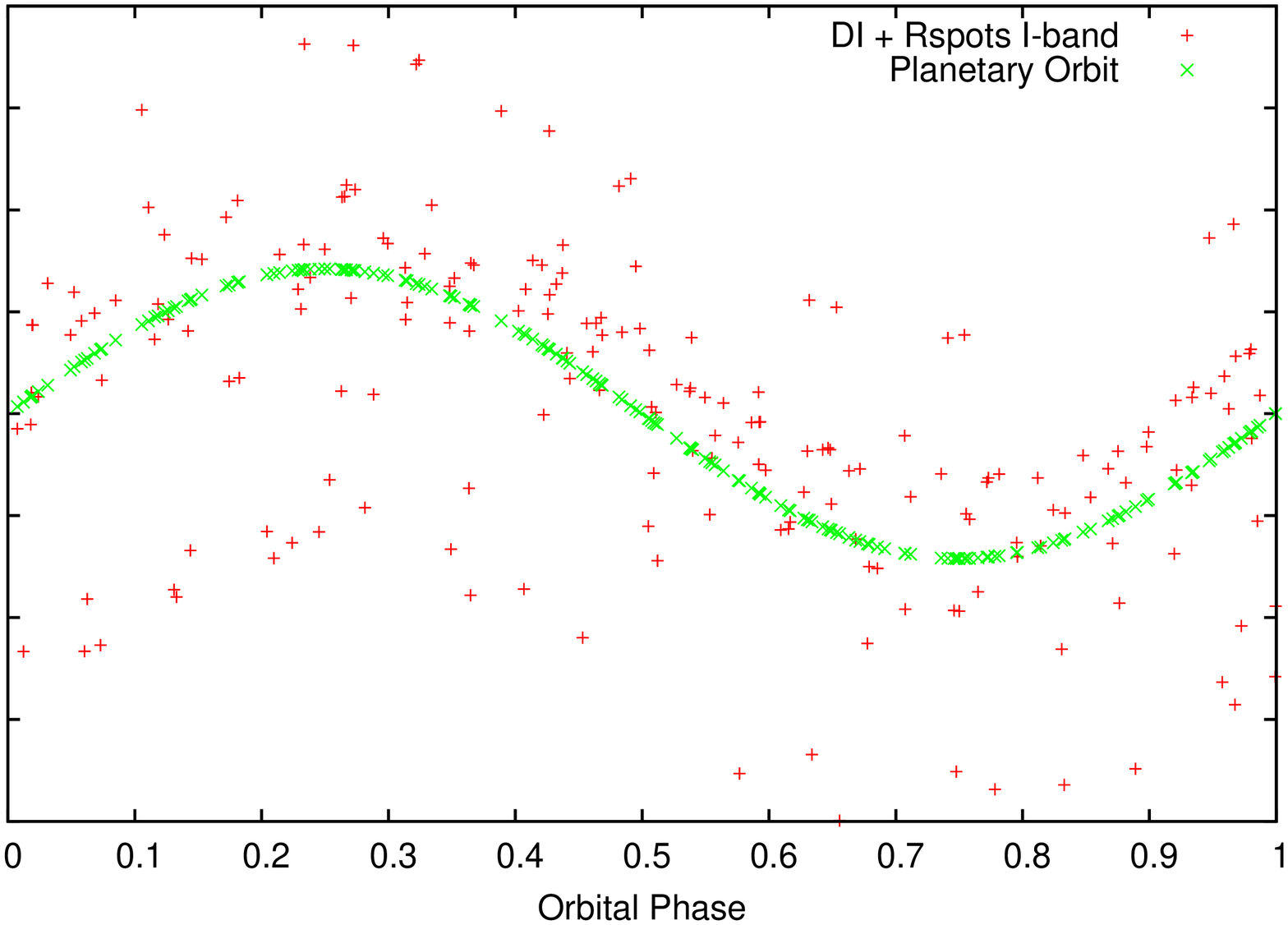}}
\subfigure[Doppler image + Plage + Random Spots I-band]{\includegraphics[angle=270,scale=0.22]{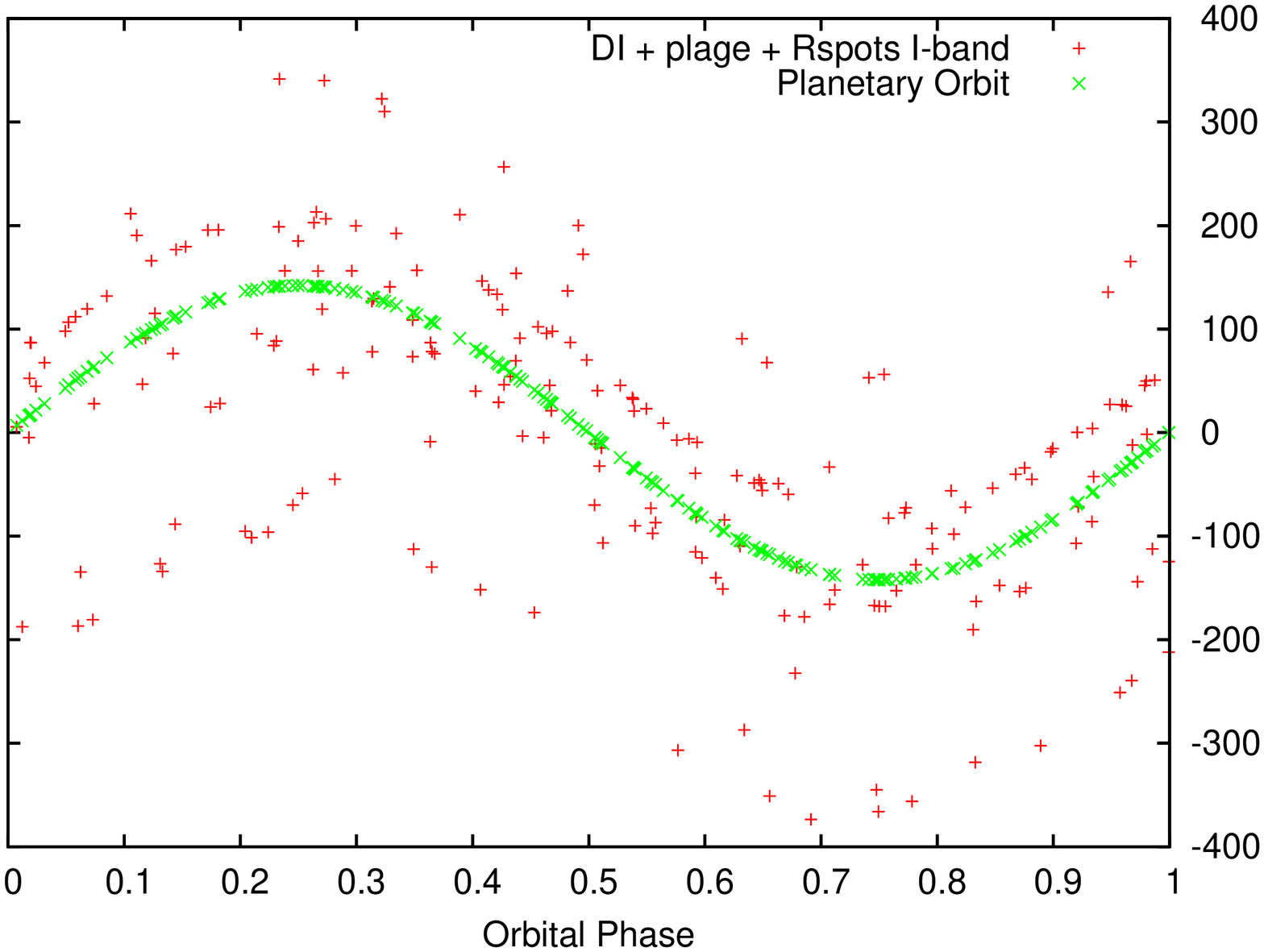}} 
\subfigure[Doppler image + Random Spots YJH-band]{\includegraphics[angle=270,scale=0.22]{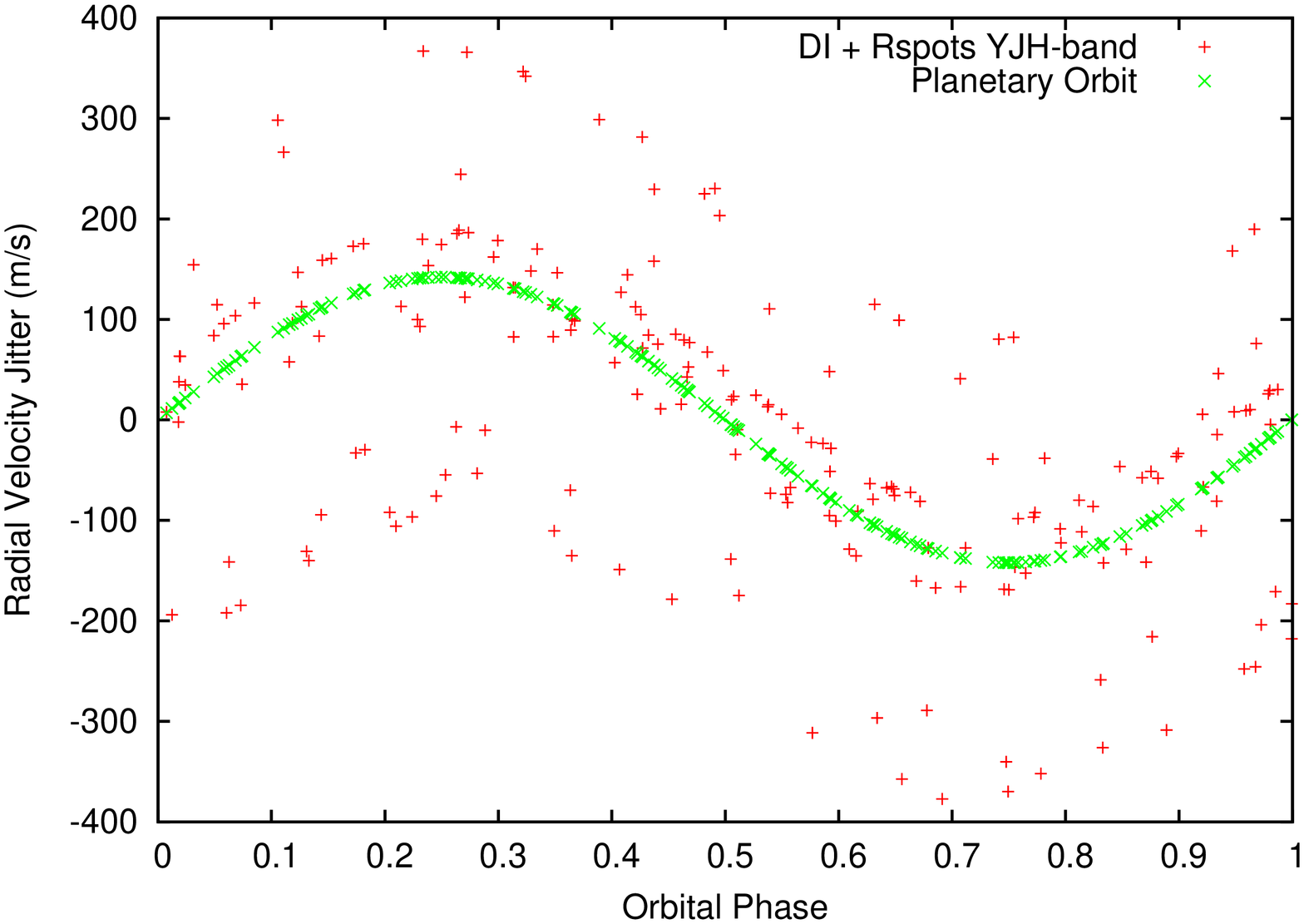}}
\subfigure[Doppler image + Plage + Random Spots YJH-band]{\includegraphics[angle=270,scale=0.22]{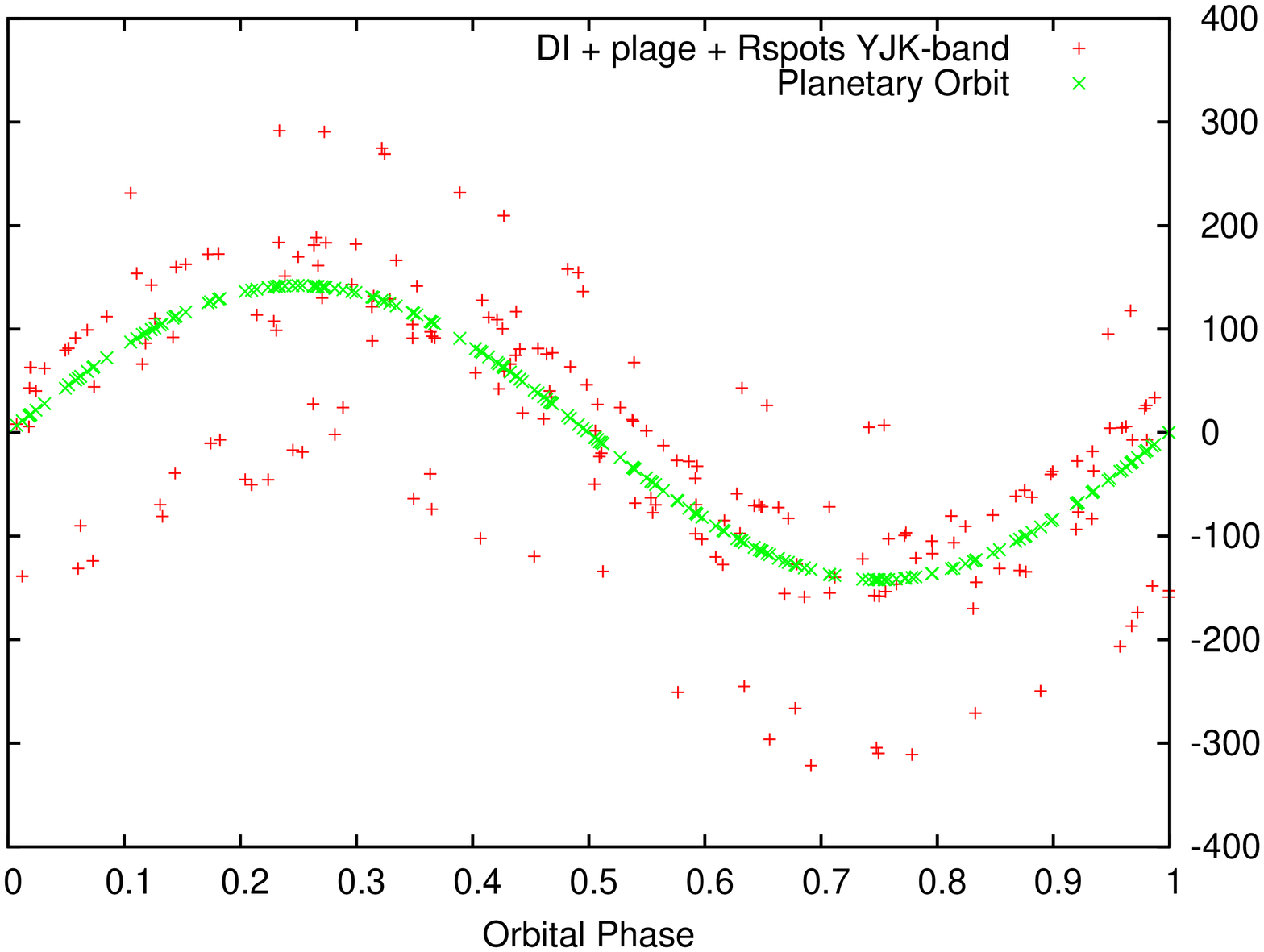}} \\
\caption{The impact of stellar activity on the simulated radial velocity profile, with the planetary radial velocity curve plotted in green and the stellar activity jitter plotted as red points.  The models shown here are for a planet mass = 10 M$_{\oplus}$ orbiting the G dwarf at 0.01 AU.}  
\protect\label{f-jitter-RV} 
\end{center}
\end{figure*}


\begin{table*}
\begin{tabular}{l c c c c c c c c c c c c }
\hline 
K dwarf & \multicolumn{12}{c}{M$_{Planet}$ M$_{\oplus}$ (M$_{\mathrm{J}}$)} \\
a (AU): & 1 & 2 & 5 & 10 & 20 & 50 & 100 & 159 (0.5) & 318 (1.0) & 636 (2.0) & 1589 (5.0) & Period (d)\\
\hline
0.01 & 1.00 & 2.00 & 5.00 & 9.99 & 19.98 & 49.95 & 99.91 & 158.85 & 317.71 & 794.26 & 1587.53 & 0.4082 \\
0.02 & 0.71 & 1.41 & 3.53 & 7.06 & 14.13 & 35.32 & 70.65 & 112.33 & 224.65 & 561.63 & 1122.55 & 1.1546 \\
0.05 & 0.45 & 0.89 & 2.23 & 4.47 & 8.94 & 22.34 & 44.68 & 71.04 & 142.08 & 355.20 & 709.96 & 4.5639 \\
0.1 & 0.32 & 0.63 & 1.58 & 3.16 & 6.32 & 15.80 & 31.59 & 50.23 & 100.47 & 251.17 & 502.02 & 12.9087 \\
0.2 & 0.22 & 0.45 & 1.12 & 2.23 & 4.47 & 11.17 & 22.34 & 35.52 & 71.04 & 177.60 & 354.98 & 36.5112 \\
0.5 & 0.14 & 0.28 & 0.71 & 1.41 & 2.83 & 7.06 & 14.13 & 22.47 & 44.93 & 112.33 & 224.51 & 144.3232 \\
1.0 & 0.10 & 0.20 & 0.50 & 1.00 & 2.00 & 5.00 & 9.99 & 15.89 & 31.77 & 79.43 & 158.75 & 408.2076 \\
2.0 & 0.07 & 0.14 & 0.35 & 0.71 & 1.41 & 3.53 & 7.06 & 11.23 & 22.47 & 56.16 & 112.26 & 1154.5853 \\
5.0 & 0.04 & 0.09 & 0.22 & 0.45 & 0.89 & 2.23 & 4.47 & 7.10 & 14.21 & 35.52 & 71.00 & 4563.8994 \\
\hline
\protect\label{t-K_kdwarf}
\end{tabular}

\begin{tabular}{l c c c c c c c c c c c c }
\hline 
G dwarf & \multicolumn{12}{c}{M$_{Planet}$ M$_{\oplus}$ (M$_{\mathrm{J}}$)} \\
a (AU): & 1 & 2 & 5 & 10 & 20 & 50 & 100 & 159 (0.5) & 318 (1.0) & 636 (2.0) & 1589 (5.0) & Period (d)\\
\hline
0.01 & 0.89 & 1.79 & 4.47 & 8.94 & 17.87 & 44.68 & 89.36 & 142.08 & 284.16 & 710.41 & 1419.93 & 0.3651 \\
0.02 & 0.63 & 1.26 & 3.16 & 6.32 & 12.64 & 31.59 & 63.19 & 100.47 & 200.93 & 502.33 & 1004.04 & 1.0327 \\
0.05 & 0.40 & 0.80 & 2.00 & 4.00 & 7.99 & 19.98 & 39.96 & 63.54 & 127.08 & 317.70 & 635.01 & 4.0821 \\
0.1 & 0.28 & 0.57 & 1.41 & 2.83 & 5.65 & 14.13 & 28.26 & 44.93 & 89.86 & 224.65 & 449.02 & 11.5459 \\
0.2 & 0.20 & 0.40 & 1.00 & 2.00 & 4.00 & 9.99 & 19.98 & 31.77 & 63.54 & 158.85 & 317.51 & 32.6566 \\
0.5 & 0.13 & 0.25 & 0.63 & 1.26 & 2.53 & 6.32 & 12.64 & 20.09 & 40.19 & 100.47 & 200.81 & 129.0866 \\
1.0 & 0.09 & 0.18 & 0.45 & 0.89 & 1.79 & 4.47 & 8.94 & 14.21 & 28.42 & 71.04 & 141.99 & 365.1120 \\
2.0 & 0.06 & 0.13 & 0.32 & 0.63 & 1.26 & 3.16 & 6.32 & 10.05 & 20.09 & 50.23 & 100.40 & 1032.6926 \\
5.0 & 0.04 & 0.08 & 0.20 & 0.40 & 0.80 & 2.00 & 4.00 & 6.35 & 12.71 & 31.77 & 63.50 & 4082.0759 \\
\hline
\end{tabular}
\caption{Simulated stellar radial velocity amplitudes induced {\em{only}} by orbiting planets from 0.01 AU to 5.0 AU with planetary masses from 1 M$_{\oplus}$ to 1589 M$_{\oplus}$ (5 M$_{\mathrm{J}}$).  The upper table shows the induced radial velocity values for the K dwarf (Mass = 0.8 M$_\odot$) and the lower table for the G dwarf (Mass = 1.0 M$_\odot$)}
\protect\label{t-KG_gdwarf}
\end{table*}


\subsection{Instrumental precision}

\begin{figure} 
\includegraphics[scale=0.35, angle=270]{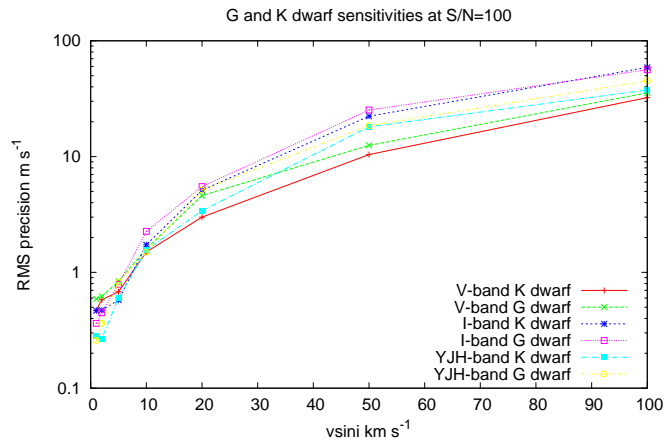}
\caption{Plot showing the achievable instrumental precision as a function of stellar \vsinis}  
\protect\label{f-insprec} 
\end{figure}

The values for the instrumental precision were computed for G and K dwarfs with a Signal-to-Noise ratio of 100.  The values are estimated by first simulating synthetic spectra over each waveband using models derived from Kurucz model atmospheres using the {\sc synthesis} code \citep{Gray1994}. The spectra were broadened to the instrumental resolution and \vsinis before adding noise (at S/N = 100). Cross-correlation with a noiseless template was then carried out a large number of times for the same S/N, but with a different random number seed. The standard deviation of the measured cross-correlation shifts was taken as the instrumental precision for each \vsinis value in each waveband.  The resulting plots are shown in Figure~\ref{f-insprec}.  From this plot it is clear that the minimum achievable precision scales directly with stellar \vsinis for all wavelength bands.  This consistent with the results of \cite{Reiners2010} for M dwarfs.  Figure~\ref{f-insprec} shows that at all \vsinis values and wavelength bands, the RMS jitter is higher for G dwarfs than for K dwarfs, while the minimum achievable instrumental precision is lowest in the V-band models and highest in the I-band models.  These values were computed with an instrumental resolution of 70,000.  Increasing the spectral resolution will have the impact of lowering these values. 

\subsection{Final models}

For each of the above star/planet combinations for G and K dwarfs we generate radial velocity profiles for a range of observation epochs corresponding to 10, 20, 50, 100 and 200 nights.  We simulate one observation epoch per night, with a random observational phase, and on consecutive nights. For each star-planet model, jitter from two sources (i.e. starspots and instrumental) are then added to the planetary radial velocities.  

As an example, the resulting radial velocity curves are plotted for a planet mass = 10 M$_{\oplus}$ orbiting the G dwarf at 0.01 AU are summarised in Figure~\ref{f-jitter-RV} for each of the stellar activity models.  The amplitude of the combined radial velocity profile is strongly dependent on the stellar activity model and directly correlates with the rms values for the G dwarf shown in Figure~\ref{f-rms-maxjitter}.  The addition of stellar activity jitter can clearly increase the radial velocity signature by up to approximately 200 m s$^{-1}$.

\section[]{Planet detection}

\begin{figure*}
\begin{center}
\subfigure{\includegraphics[angle=270,scale=0.32]{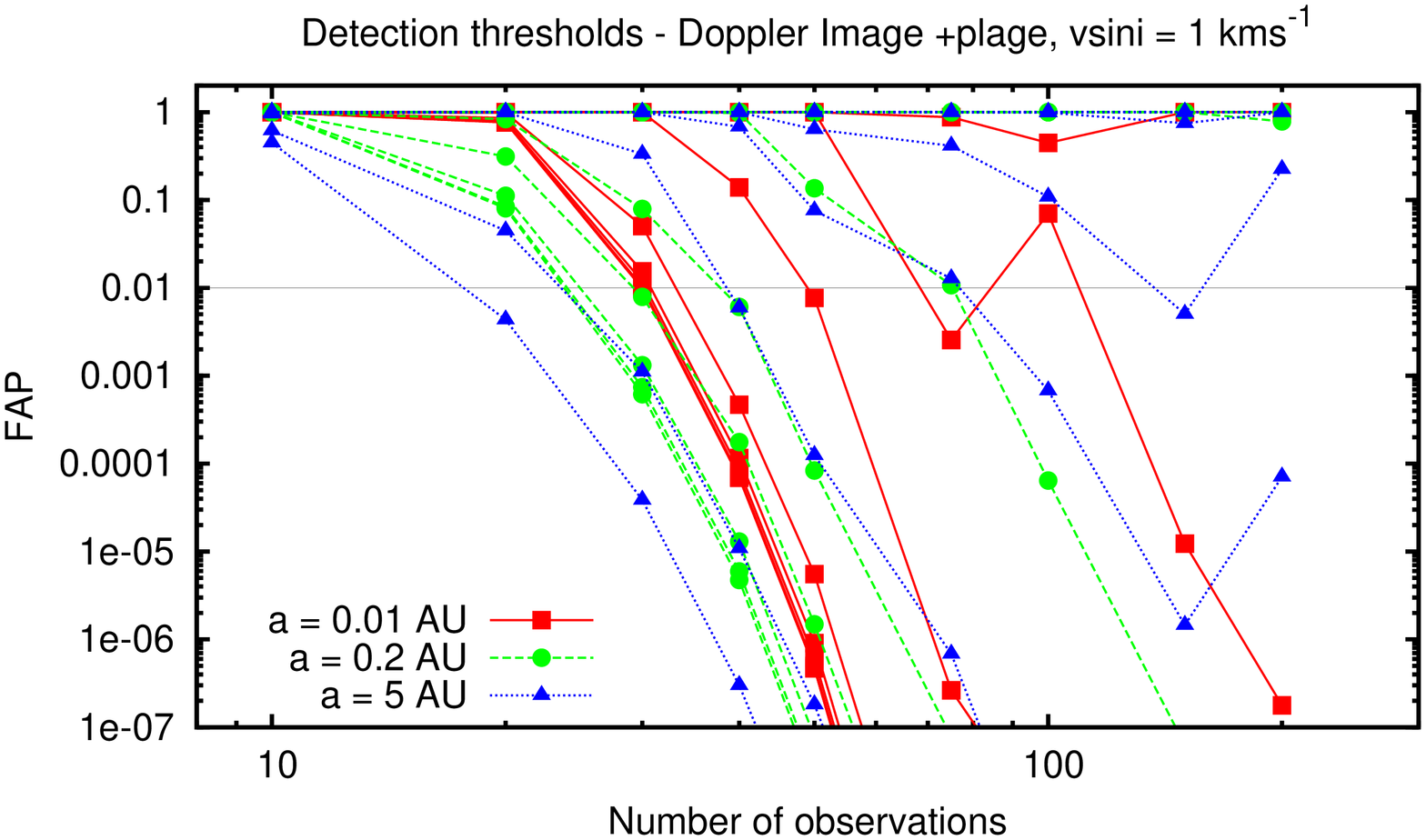}} 
\subfigure{\includegraphics[angle=270,scale=0.32]{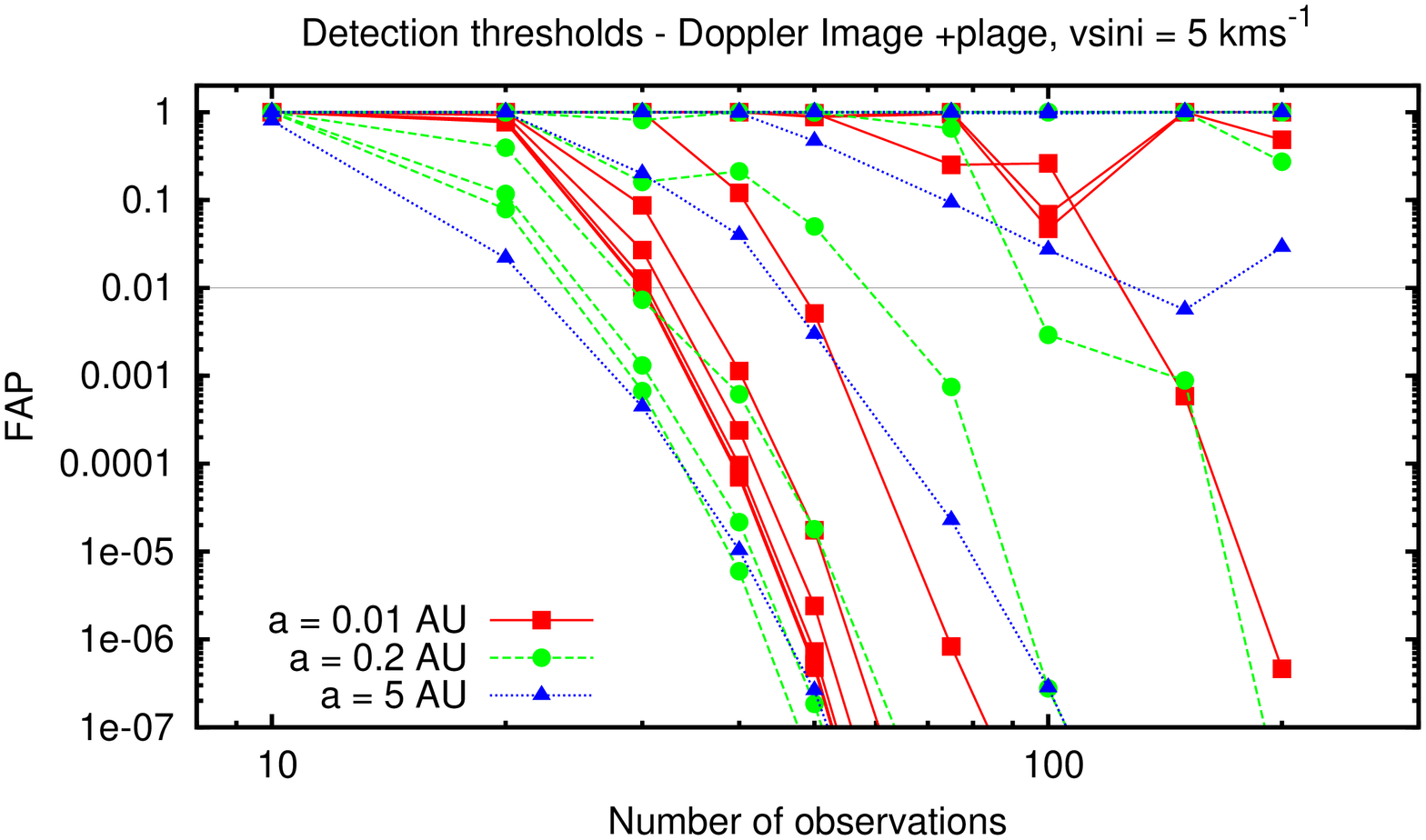}} \\
\subfigure{\includegraphics[angle=270,scale=0.32]{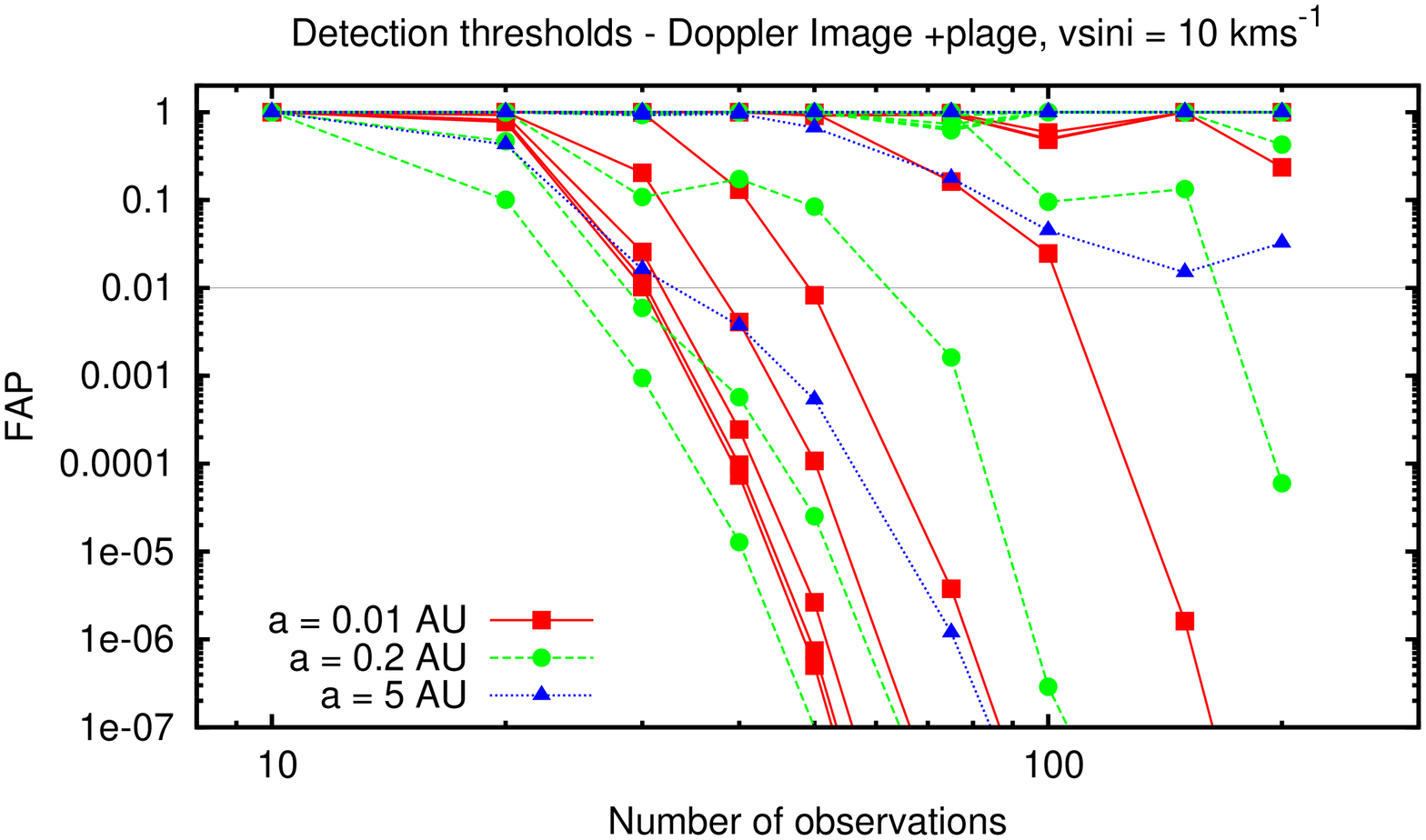}}
\subfigure{\includegraphics[angle=270,scale=0.32]{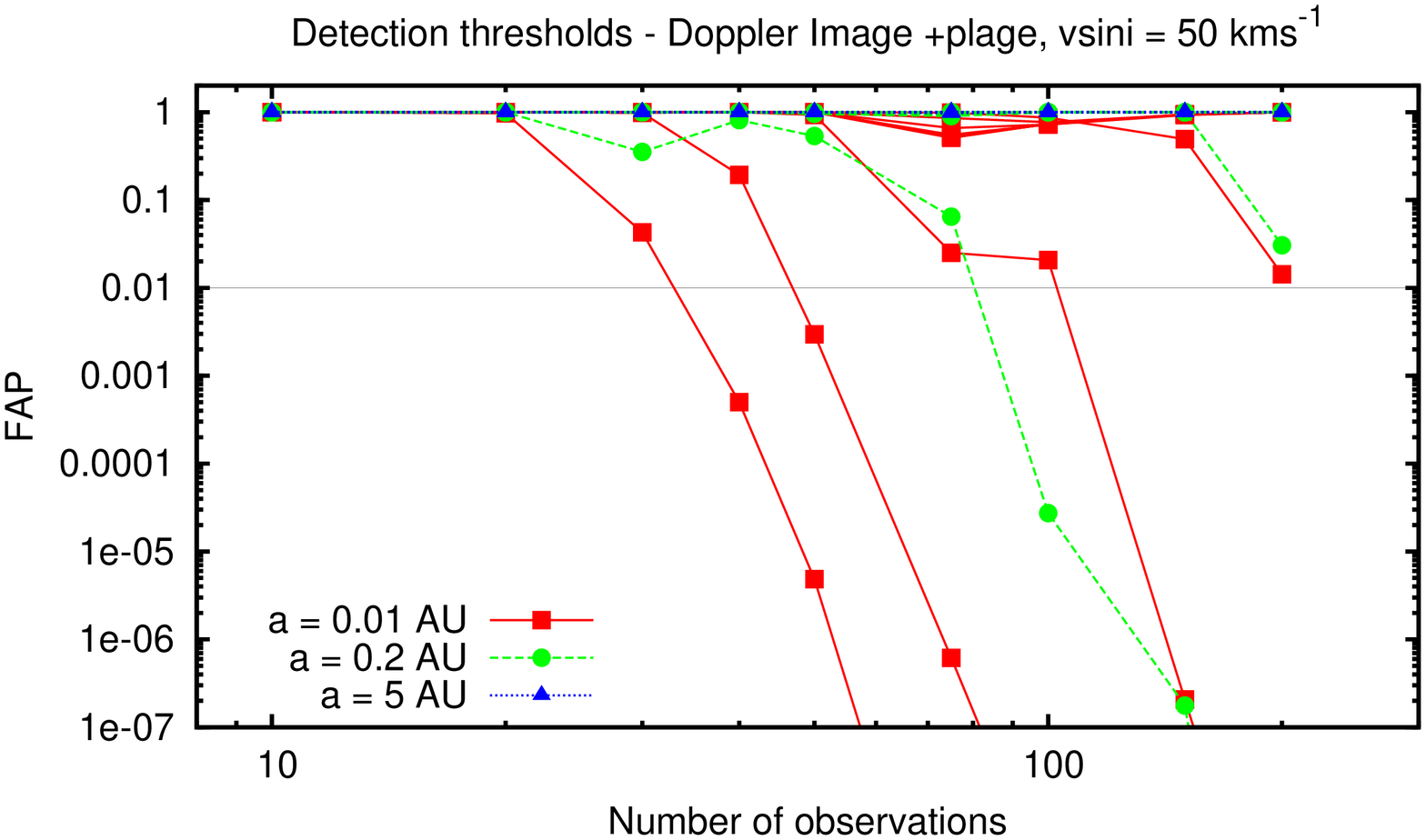}} \\
\caption{Detection false alarm probabilities (FAPs) as a function of number of observations epochs for 5, 10, 20, 50, 100, 159, 318, 636 and 1589 M$_{\oplus}$ planet orbiting at 0.01 AU, 0.2 AU and 5 AU [red, green and blue lines in plots].  The stellar activity model used for these plots is the model containing Doppler image and additional plage.  The horizontal line indicates the 1 per cent false alarm probability (FAP=0.01) with the region above this line representing undetected planets.   All points with FAP $<0.01$ are considered as planet detections.  The top left plot is with \vsinis = 1 km s$^{-1}$, top right with \vsinis = 5 km s$^{-1}$, bottom left with \vsinis = 10 km s$^{-1}$, and bottom right with \vsinis = 50 km s$^{-1}$.}  
\protect\label{f-spider-di_plage} 
\end{center}
\end{figure*}


\begin{figure*} 
\begin{center}
\subfigure[Doppler image (DI)]{\includegraphics[angle=270,scale=0.3]{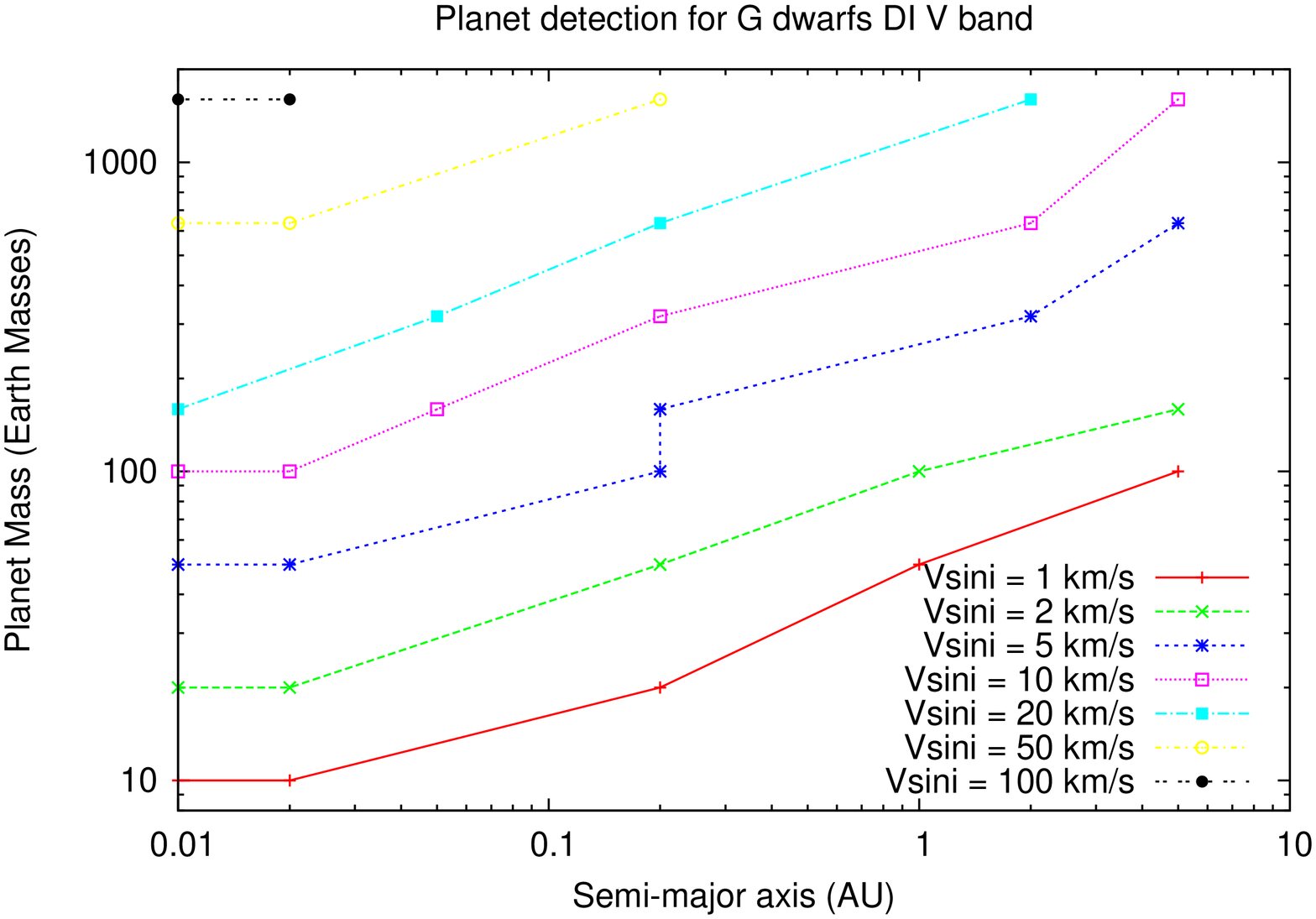}}
\subfigure[DI + Plage]{\includegraphics[angle=270,scale=0.3]{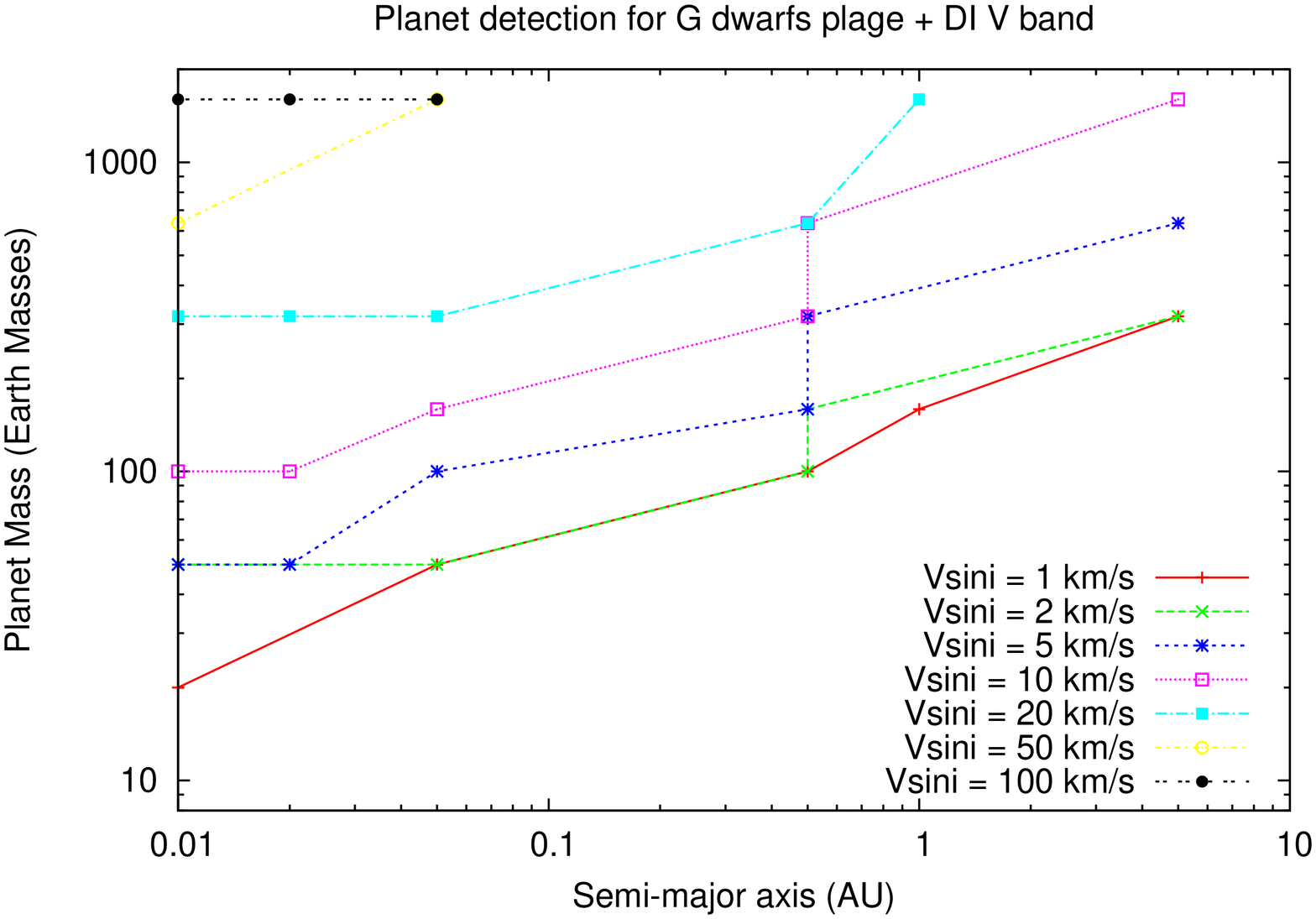}} \\
\subfigure[DI + Random Spots]{\includegraphics[angle=270,scale=0.3]{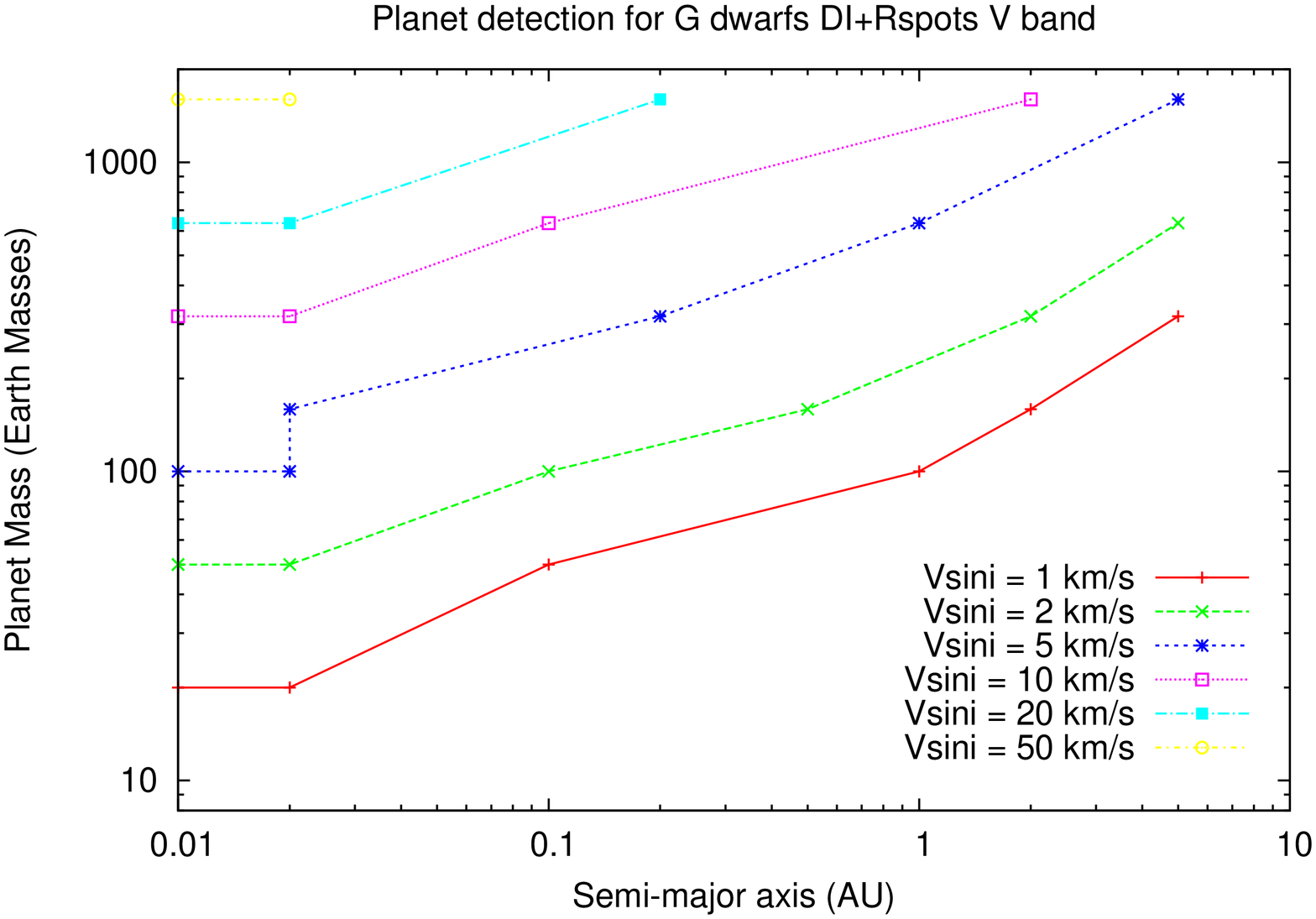}}
\subfigure[DI + plage + Random Spots]{\includegraphics[angle=270,scale=0.3]{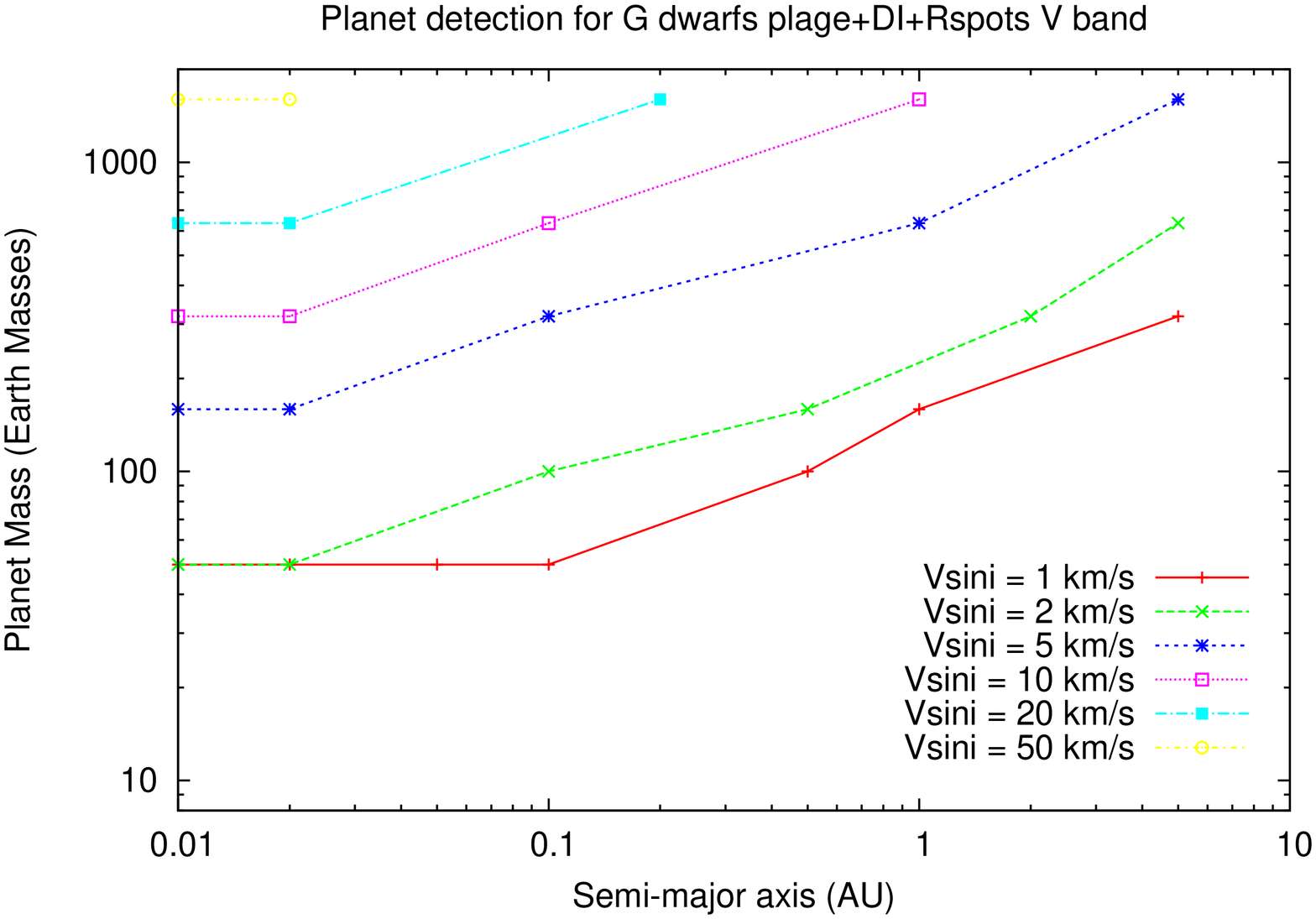}}\\
\caption{Planet detection as a function of planetary mass and star - planet separation.  Each point is the lowest mass planet detected within 50 observational epochs in Figure~\ref{f-spider-di_plage} for a given star planet separation.  The connected points represent a lower limit of detections for a given stellar \vsinis value with planets detected in the area to the left of each connecting line (per \vsinis value).  The stellar activity models used are the Doppler image (top left), Doppler image + plage (top right), Doppler image + Rspots (bottom left), with a temperature contrast of T$_{sp}$=0.65 T$_{phot}$. The step-changes in the plotted data are a result of small number statistics, preventing a smooth trend when plotted on a log-scale.} 
\protect\label{f-planetdetection-dip} 
\end{center}
\end{figure*}


\begin{figure*} 
\begin{center}
\subfigure[G dwarf minimum planet detection at 0.05 AU]{\includegraphics[angle=270,scale=0.3]{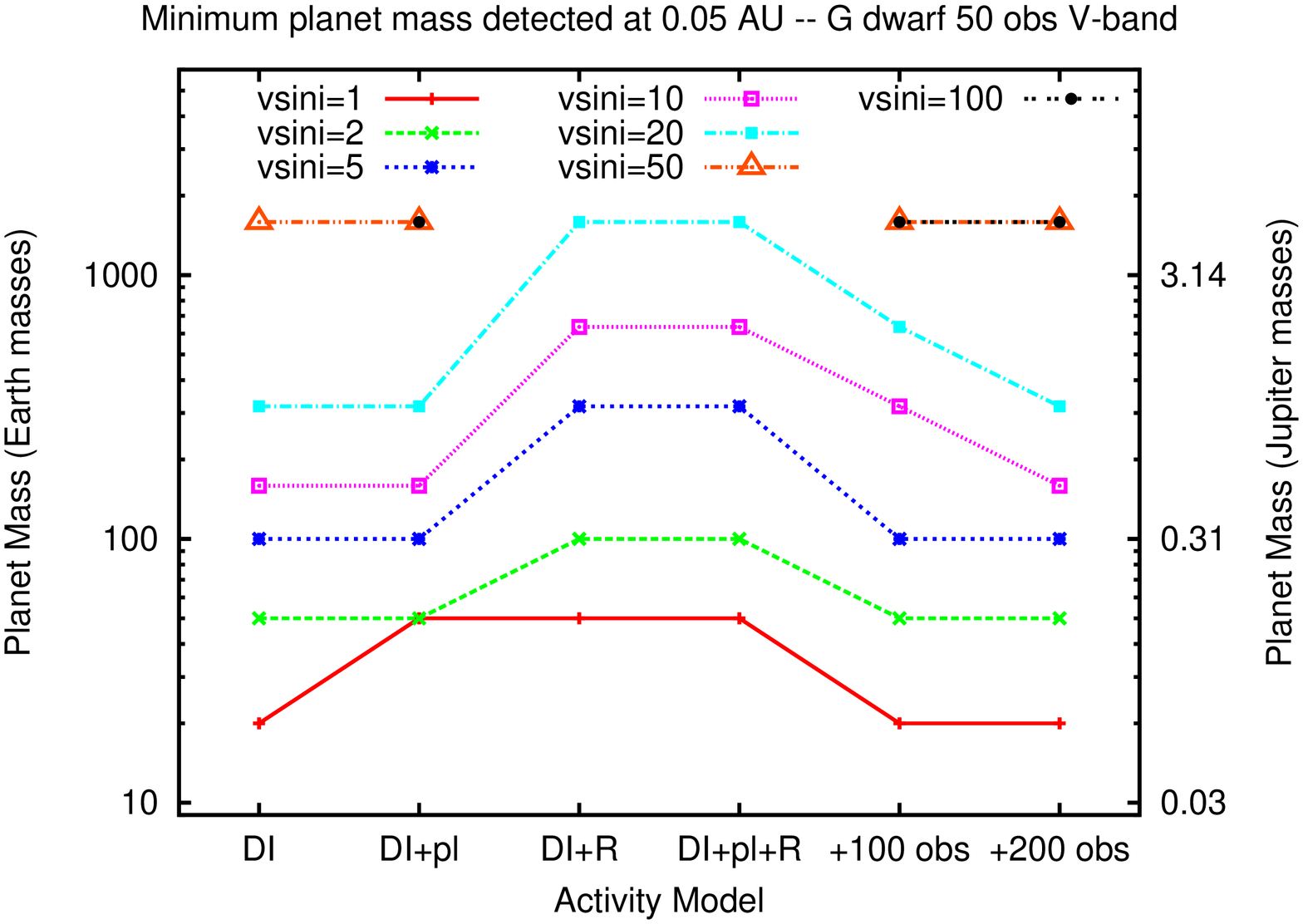}}
\subfigure[G dwarf minimum planet detection at 1 AU]{\includegraphics[angle=270,scale=0.3]{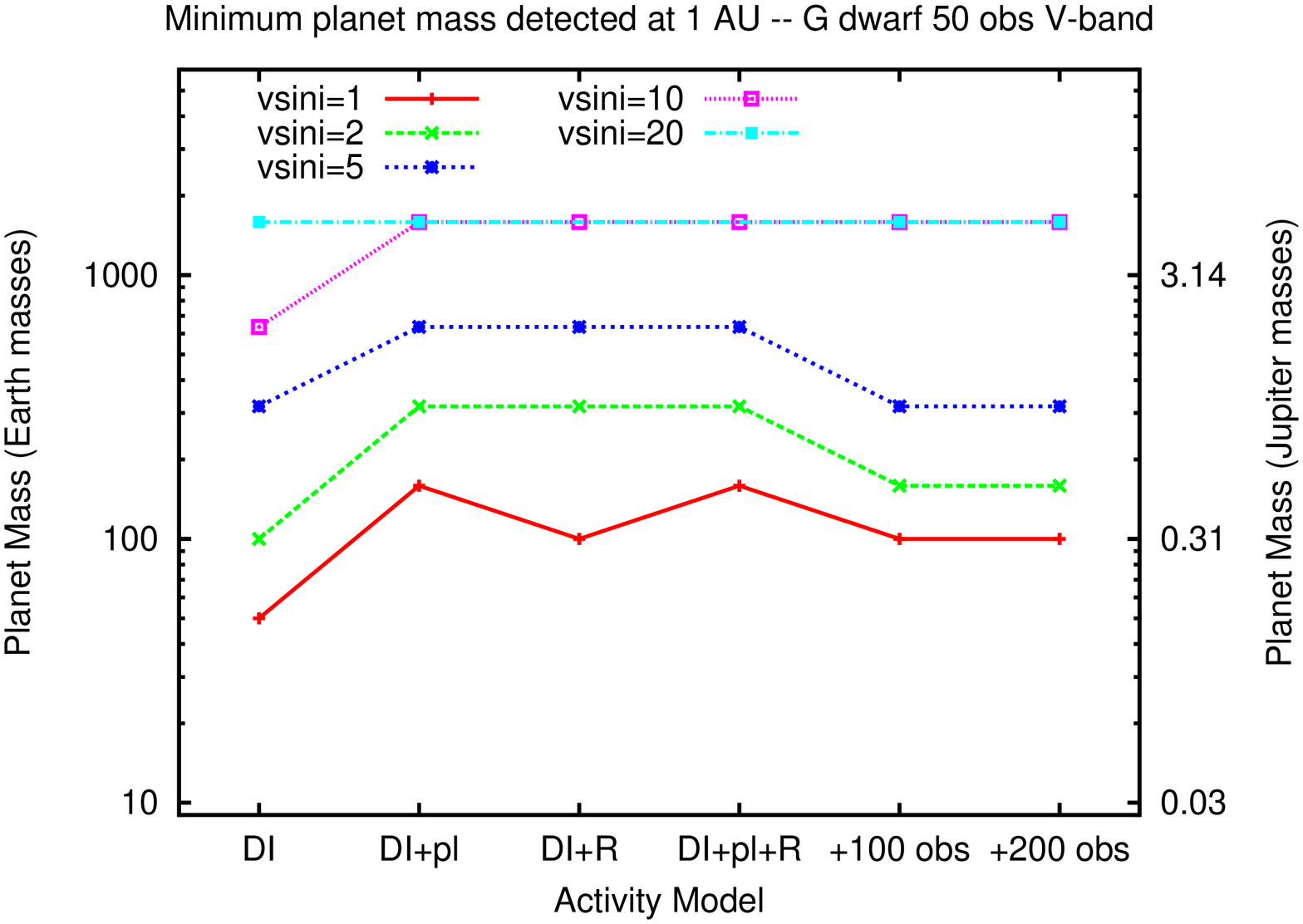}} \\
\subfigure[K dwarf minimum planet detection at 0.05 AU]{\includegraphics[angle=270,scale=0.3]{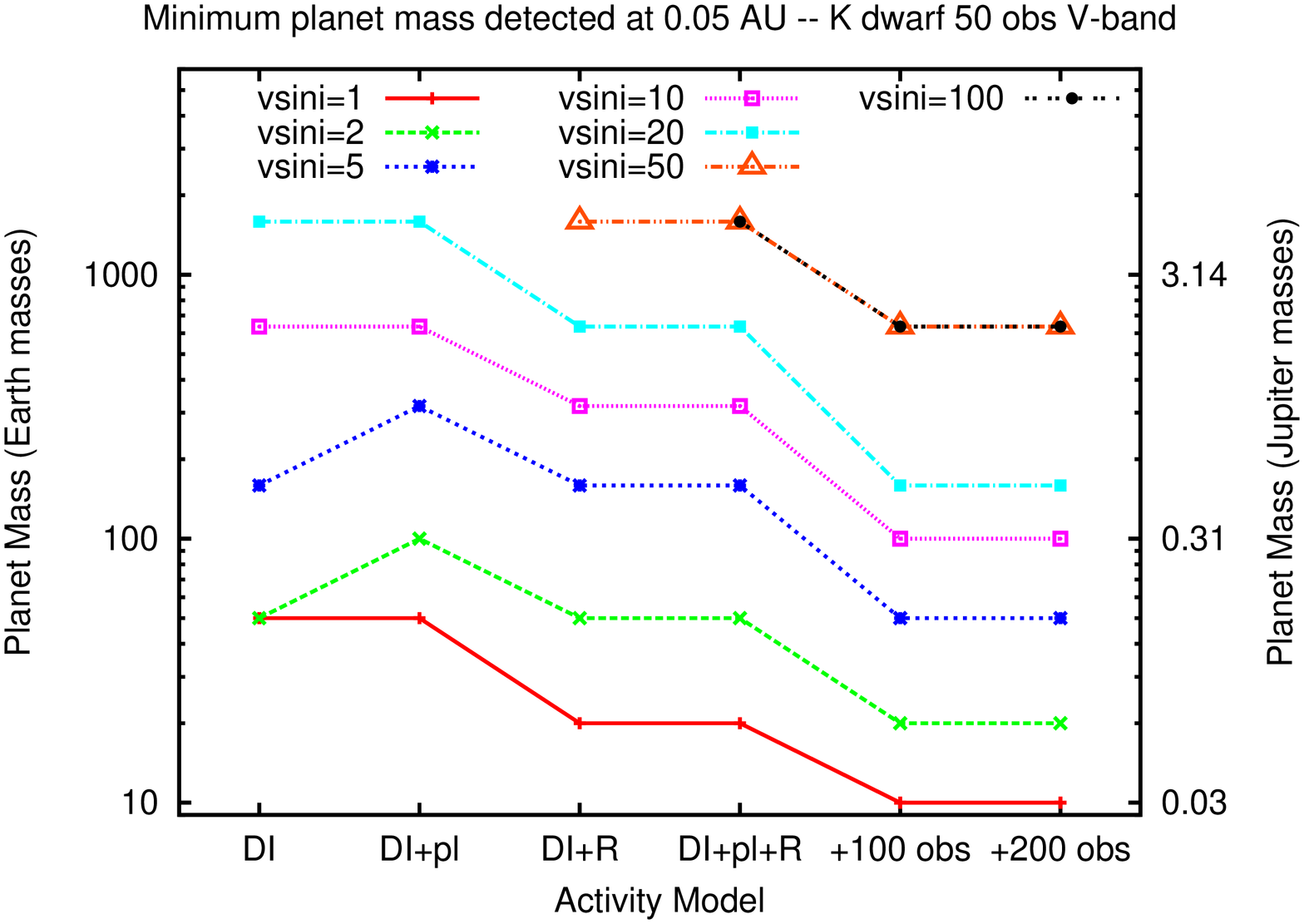}}
\subfigure[K dwarf minimum planet detection at 0.5 AU]{\includegraphics[angle=270,scale=0.3]{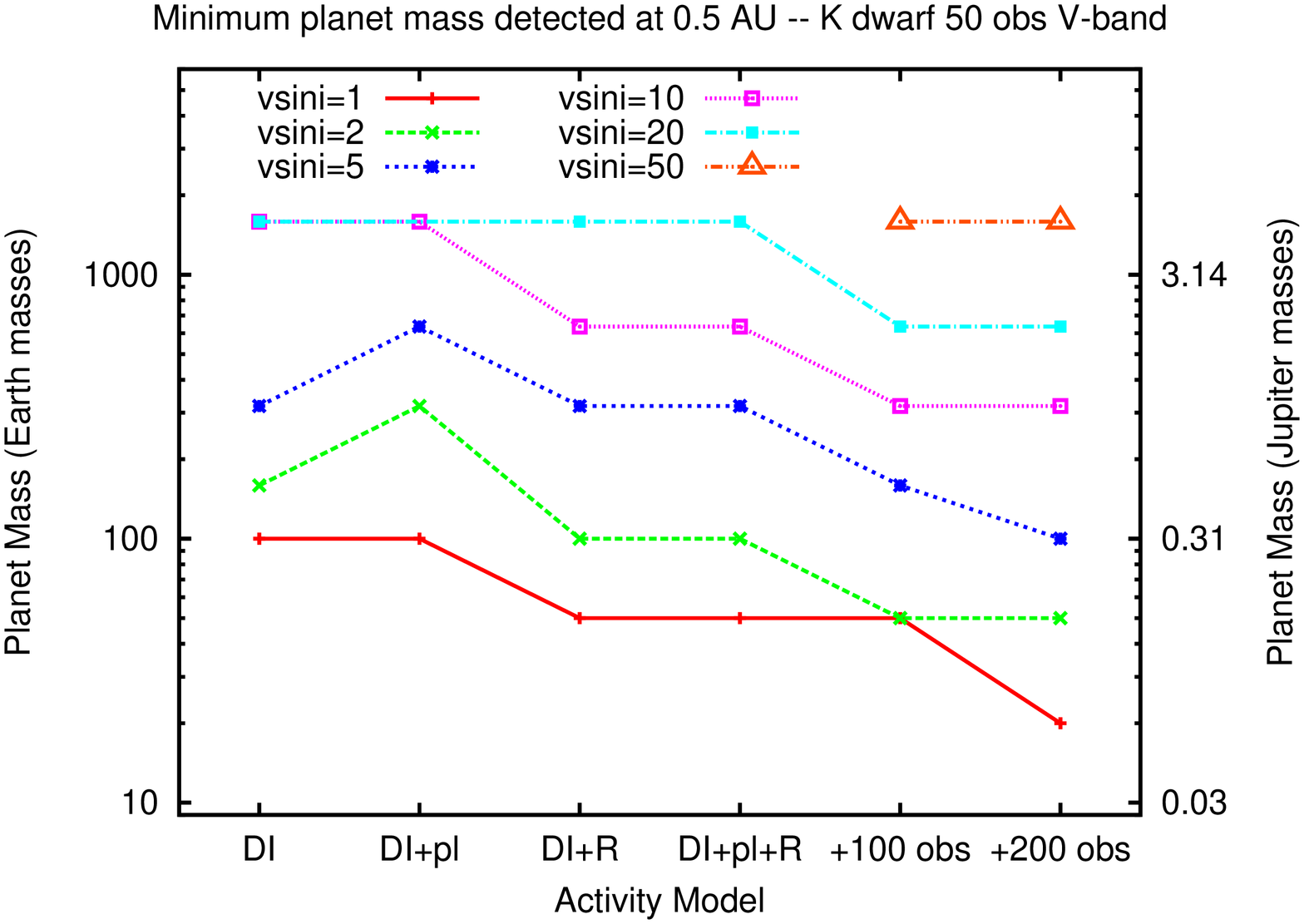}}\\
\caption{Summary of the minimum planetary mass detected within 50 observational epochs for the Stellar activity model: DI (Doppler image), DI+pl (Doppler image + plage), DI + R (Doppler image + Random Spots), DI + plage + R (Doppler image + plage + Random Spots), + 100 obs (Doppler image + plage + Random Spots with 100 observational epochs) + 200 obs (Doppler image + plage + Random Spots with 200 observational epochs).  The models are at V-band wavelengths and the maximum planetary mass modelled is 5 M$_{\mathrm{J}}$ which are binned following the masses of the simulated planets.}  
\protect\label{f-summaryplots-Vband} 
\end{center}
\end{figure*}

\begin{figure*} 
\begin{center}
\subfigure[G dwarf minimum planet detection at 0.05 AU]{\includegraphics[angle=270,scale=0.3]{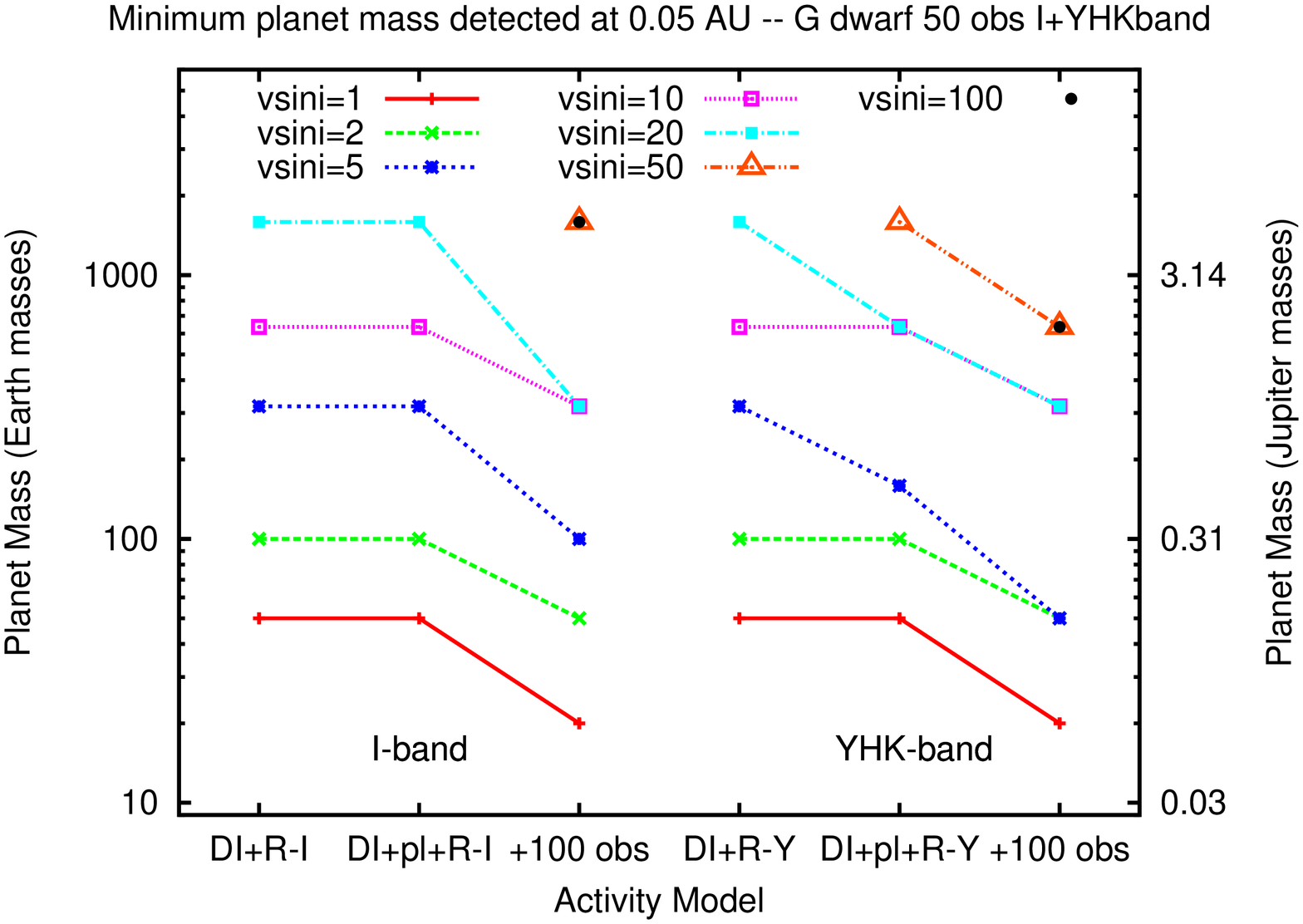}}
\subfigure[G dwarf minimum planet detection at 1 AU]{\includegraphics[angle=270,scale=0.3]{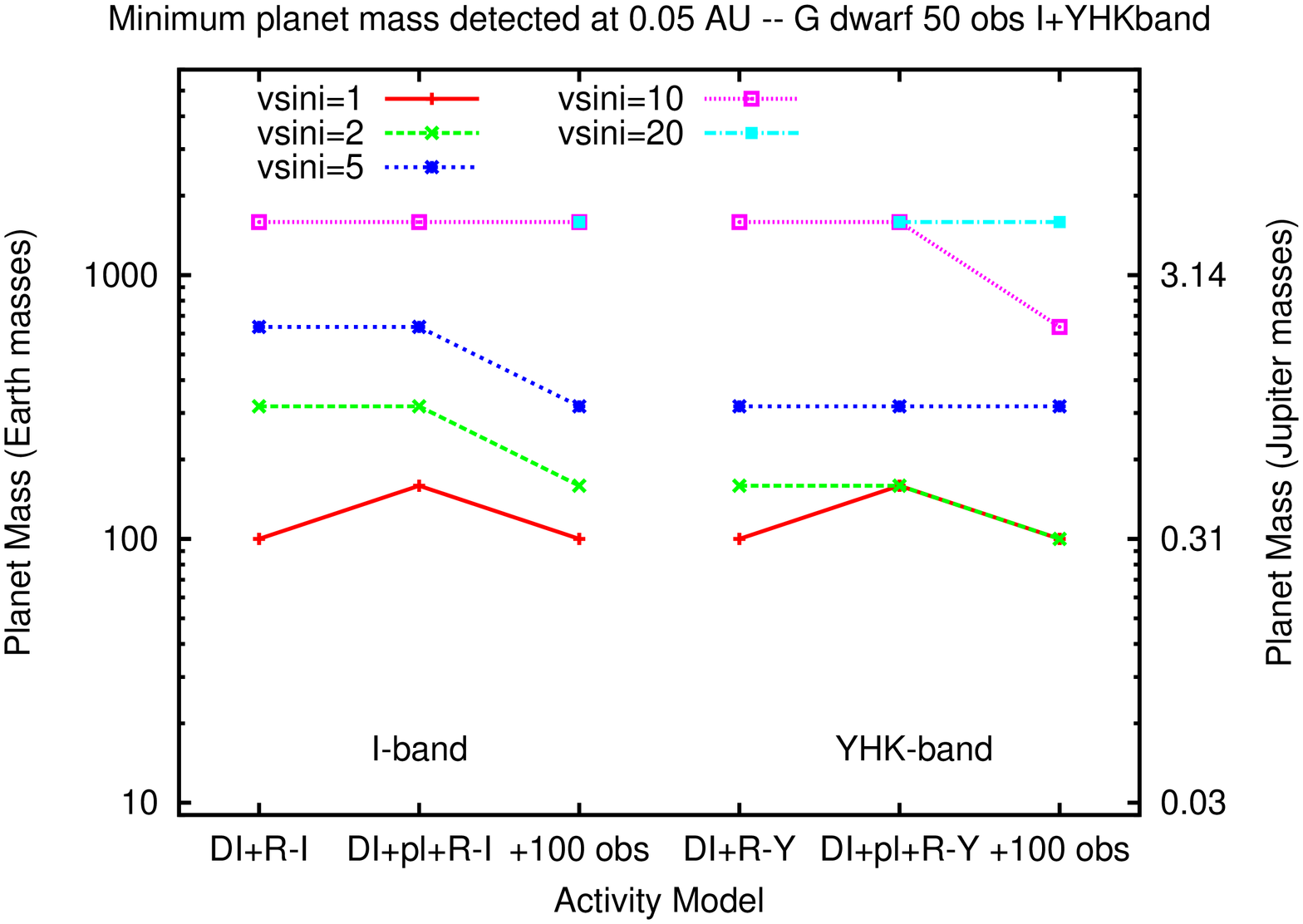}} \\
\subfigure[K dwarf minimum planet detection at 0.05 AU]{\includegraphics[angle=270,scale=0.3]{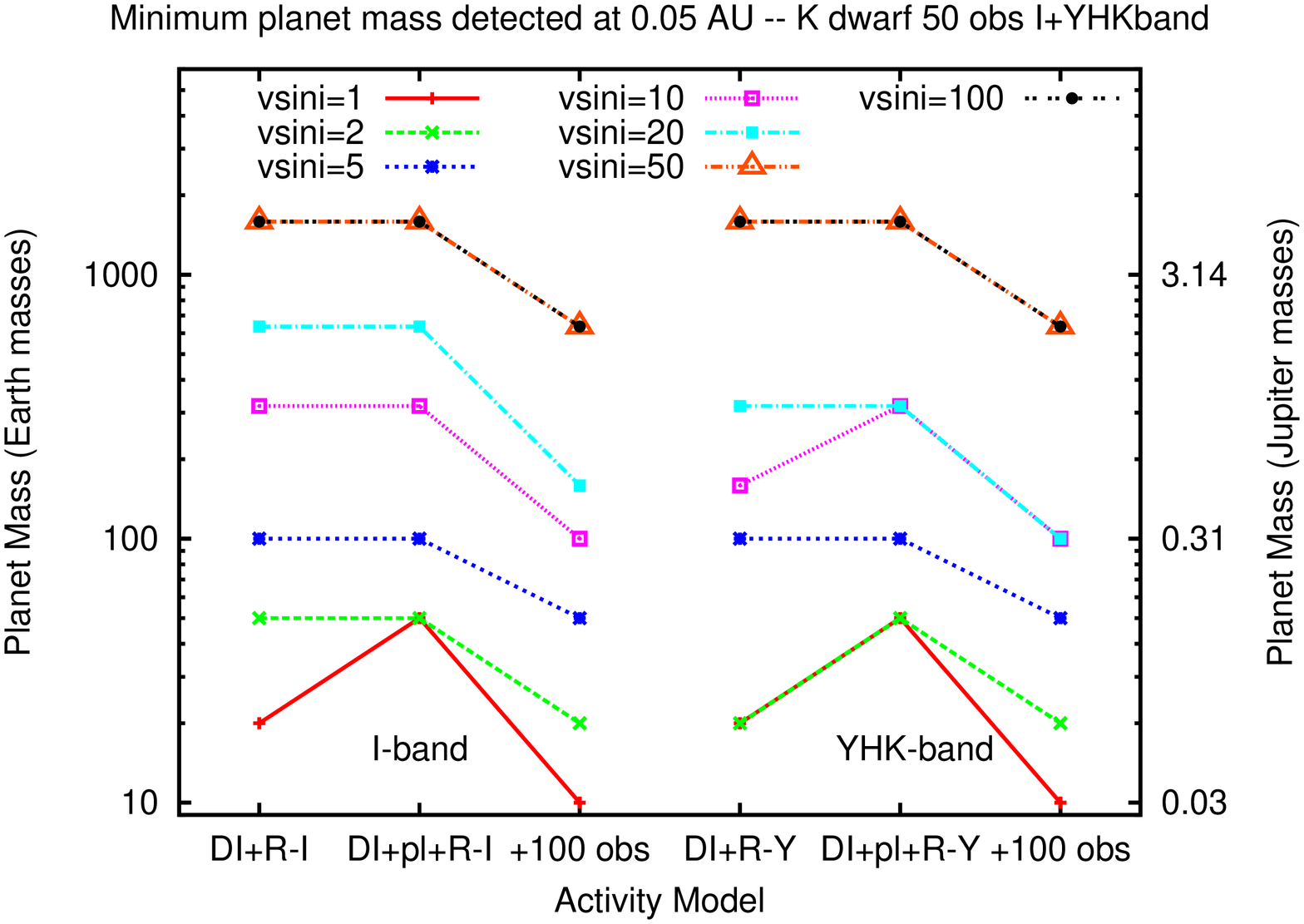}}
\subfigure[K dwarf minimum planet detection at 0.5 AU]{\includegraphics[angle=270,scale=0.3]{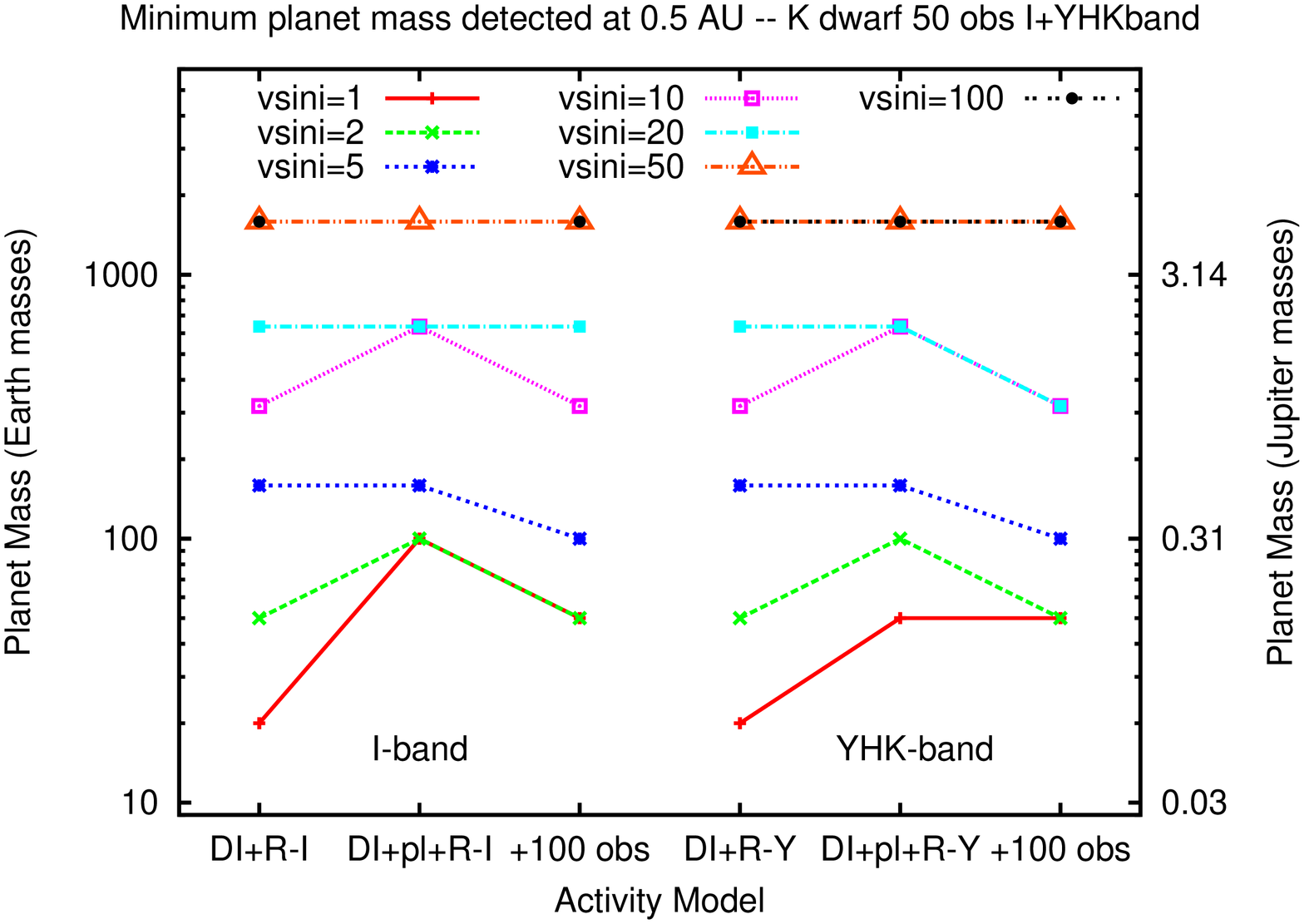}}\\
\caption{As per Figure~\ref{f-summaryplots-Vband} but for I-band (indicated by -I in the x-axis values) and YHK-band (indicated by -Y in the x-axis values).  The maximum planetary mass modelled is 5 M$_{\mathrm{J}}$ which are binned following the masses of the simulated planets.}  
\protect\label{f-summaryplots-I-YHKband} 
\end{center}
\end{figure*}

To determine realistic limitations for detecting planets around active G and K dwarfs we then apply the commonly used Lomb Scargle periodogram planet-finding algorithm \citep{Barnes2011, press1992} on each radial velocity curve for each model.  Highlights of the resulting periodogram analysis are presented in Figure~\ref{f-spider-di_plage} as detection thresholds.  

In Figure~\ref{f-spider-di_plage}, the false alarm probability is plotted against the total number of observations for \vsinis values of 1, 5, 10 and 50 km s$^{-1}$ at star-planet separations of 0.01 AU, 0.2 AU and 5.0 AU for the model with a Doppler image and additional plage.  The clear detection of planets is shown for each simulation where the false alarm probability is below 0.01 (indicated by a horizontal dashed line).  Above this limit there are no claimed or candidate planet detections that satisfy the 1\% false-alarm threshold.   For the case where \vsinis = 1 kms$^{-1}$, planets orbiting at 0.01 AU and with masses down to 5 M$_{\oplus}$ are detected within 200 observing epochs or 10 M$_{\oplus}$ within 50 epochs.  Moving further out to star-planet separations of 0.2 AU, increases the minimum detectable planetary mass to 10 M$_{\oplus}$ within 200 observing epochs and 20 M$_{\oplus}$ within 50 epochs.  At the largest separation of 5 AU, the minimum detected mass is 159 M$_{\oplus}$ or 0.5 M$_{\mathrm{J}}$.

Increasing the \vsinis values to 5 kms$^{-1}$, also increases the minimum detectable planetary mass at separations of 0.01 AU to 20 M$_{\oplus}$ within 200 observing epochs or 159 M$_{\oplus}$ (0.5 M$_{\mathrm{J}}$) within 50 epochs.  At an increased star-planet separation of 0.2 AU it is possible to detect 
a 159 M$_{\oplus}$ planet within 100 observing epochs or 318 M$_{\oplus}$ (1 M$_{\mathrm{J}}$) within 50 epochs.  While at 5 AU, planet detections of 318 M$_{\oplus}$ within 200 observing epochs or a much larger planet of 636 M$_{\oplus}$ (2 M$_{\mathrm{J}}$) within 50 epochs are possible.

The trend of planet detectability decreasing with increasing stellar \vsinis continues for \vsinis values of 10 kms$^{-1}$ and 50 kms$^{-1}$.  In the latter case, 159 M$_{\oplus}$ planets can be detected close into the star at 0.01 AU with 200 epochs, while only 1589 M$_{\oplus}$ (5 M$_{\mathrm{J}}$) planets can be detected within 50 observations epochs.  At separations of 0.2 AU, only 1589 M$_{\oplus}$ planets can be detected within 100 epochs and no planets are detected further out at 2 AU.  

\subsection{Spot/photosphere temperature contrast}

We find that for G dwarfs the spot/photosphere temperature contrast ratio makes negligible difference to the detection of planets, while for K dwarfs the temperature contrast T$_{spot}$=T$_{phot}$-1500K slightly increases the probability of detecting a planet at 0.02AU but only in the case of \vsinis=20km s$^{-1}$.  For the rest of this analysis we will only discuss the results for the case where T$_{spot}$=0.65 T$_{phot}$.  

\subsection{Lower limits of planet detection}
\protect\label{s-lowerpldet}

As we showed in paper I, the ability to detect a planetary radial velocity signature decreases with increasing \vsinis and stellar activity level.  To simplify the presentation of the results for the other stellar activity models, we have summarised the planetary detections shown in Figure~\ref{f-spider-di_plage}, by plotting only the {\em minimum detected planet mass within 50 observing epochs} on a plot of Planet Mass as a function of orbital radius.  The plots are then reproduced for values of \vsinis = 1, 2, 5, 10, 20, 50 and 100 km s$^{-1}$.  The coloured lines represent the minimum detectable planets, with the area to the left of the plotted lines indicating regions with detected planets for a given stellar \vsinis.  A selection of these plots for G-dwarf models at V-band wavelengths (Doppler image, Doppler image + plage, Doppler image + Random Spots, Doppler image + plage + Random Spots) are shown in Figure~\ref{f-planetdetection-dip}. The step-changes in the plotted data are a result of small number statistics, preventing a smooth trend when plotted on a log-scale.

The models using the Doppler image stellar activity model (Figure~\ref{f-planetdetection-dip} top left panel) show the highest number of planet detections with even a 10 M$_{\oplus}$ planet being detected at \vsinis values of 1 km s$^{-1}$.  At a typical orbital separation of 0.05 AU for a Jupiter mass planet, it is possible to detect 20 M$_{\oplus}$ planets at \vsinis values of 1 km s$^{-1}$, 159 M$_{\oplus}$ (1 M$_{\mathrm{J}}$) planets at \vsinis values of 20 km s$^{-1}$ and 1589 M$_{\oplus}$ (5 M$_{\mathrm{J}}$) at more extreme rotation values of 50 km s$^{-1}$.  A typical planet orbiting in the habitable zone of a G dwarf would be detectable for planetary mass of 50 M$_{\oplus}$ at \vsinis values of 1 km s$^{-1}$, 100 M$_{\oplus}$ at 2 km s$^{-1}$, up to 1589 M$_{\oplus}$ (5 M$_{\mathrm{J}}$) at faster rotation velocities of 20 km s$^{-1}$.  

The impact of plage in the stellar activity models (Figure~\ref{f-planetdetection-dip} top right panel) is strongest for the models with lower \vsinis values.  For example, the minimum detectable planetary mass is now 50 M$_{\oplus}$ at 0.05 AU and 159 M$_{\oplus}$ (0.5 M$_{\mathrm{J}}$) at 1 AU for \vsinis values of 1 km s$^{-1}$.  For faster stellar rotation velocities i.e. $\geq$ 10 km s$^{-1}$ the difference is less prominent, though the stellar activity model with plage detects slightly less massive planets than the Doppler image activity model.  This can be caused by surface features, i.e. plage and spots, becoming better resolved at higher \vsinis values and causing these opposing distortions to be resolved in the profile in such a way that they cancel each other out.  Such subtle effects can have significant consequences and can lead to deviations from the expected trends.

The addition of many random spots to the Doppler image stellar activity model (Figure~\ref{f-planetdetection-dip} bottom left panel) considerably increases the minimum detectable planetary mass at all stellar \vsinis values, with no planets being detected at the extreme \vsinis values of 100 km s$^{-1}$, even for close-in planets.  At 0.05 AU, the minimum detectable planet is now 50 M$_{\oplus}$ at \vsinis = 1 km s$^{-1}$ up to a maximum of 1589 M$_{\oplus}$ (5 M$_{\mathrm{J}}$) at \vsinis values of only 20 km s$^{-1}$.  At larger separations of 1 AU this decreases to \vsinis values of 10 km s$^{-1}$ to detect a planet of 1589 M$_{\oplus}$ (5 M$_{\mathrm{J}}$).  The addition of plage (Figure~\ref{f-planetdetection-dip} bottom right panel) again shows the strongest impact at low stellar \vsinis values i.e. at 1 and 2 km s$^{-1}$.  For higher \vsinis values the impact of the plage is almost negligible.

\subsection{Planet Detection Thresholds}

For each stellar activity model and for G and K dwarfs, the minimum detected planet thresholds within 50 observational epochs are summarised in Figure~\ref{f-summaryplots-Vband} for V-band models and Figure~\ref{f-summaryplots-I-YHKband} for I-band and YHK-band models for planets detected at 0.05 AU (G and K dwarfs) and 0.5 AU (K dwarfs) and 1 AU (G dwarfs).  The star-planet separation of 0.05 AU is chosen as it is where hot Jupiter mass planets are typically found, while 1AU is in the habitable zone of the G dwarf and 0.5 AU in the habitable zone of the K dwarf.  The maximum planetary mass modelled is 5 M$_{\mathrm{J}}$ which are binned following the masses of the simulated planets.

\subsubsection{V-band}

The minimum planet masses detected at 0.05 AU and 1 AU for G dwarfs are respectively shown in the left and right top panels of Figure~\ref{f-summaryplots-Vband}.  The impact of all the stellar activity models is noticeably stronger for the detection of planets at 1 AU at all \vsinis values because the relative amplitude of the planet is smaller and can be easily hidden by the stellar activity jitter.  For close-in planets, the addition of stellar plage (model DI+p) has only a marginal impact and actually increases the detection threshold of a 5 M$_{\mathrm{J}}$ planet for the model with \vsinis = 100 km s$^{-1}$.  The increase of the number of observing epochs to 100 (for \vsinis $\leq$ 5 km s$^{-1}$,  50 km s$^{-1}$ and 100 km s$^{-1}$) compensates for the increase in stellar activity caused by the addition of many random starspots (model DI+R and DI+pl+R).  For other \vsinis values it is necessary to increase the number of observational epochs to 200 to negate the increase in stellar activity by adding random spots to the model using only the Doppler image.  For planets orbiting at 1AU (Figure~\ref{f-summaryplots-Vband}, top right panel) it is only possible to detect planets around stars that are rotating $\leq$ 20 km s$^{-1}$. The addition of plage to the Doppler image is more noticeable at all \vsinis models than for the close-in planets, but makes little difference when added to the random spot model.  The increase of the number of observing epochs to 100 increases the detection of smaller planets but does not negate the increase in stellar activity as found for the planets orbiting close-in at 0.05 AU.  There is no difference in planet detectability by increasing the number of observing epochs from 100 to 200 for all \vsinis values.  

The results summarising the minimum planetary masses detectable for planets orbiting K dwarfs at 0.05 AU and 0.5 AU are shown in the bottom two panels of Figure~\ref{f-summaryplots-Vband}.  The impact of the stellar activity model behaves quite differently compared to the results for the G dwarfs.  The addition of plage to the Doppler image activity model, does not make any difference, while the addition of many random spots to the Doppler image activity model actually results in the detection of lower mass planets.  As noted in Section~\ref{s-lowerpldet}  this can be explained by plage and spots creating opposite distortions in the absorption line profile, with the magnitude of this effect strongly depending on the input starspot pattern.  For example, the reconstructed Doppler image for AB Dor (K dwarf) has a much more uniform spot pattern than the Doppler images of HD 141943, where the starspots are more concentrated in large spot groups.  The additional plage also reflects this distribution resulting in the maximum activity induced starspot jitter.  However, when the additional random spots are added to this image it dilutes the impact of the plage for the K dwarf by making the profiles more symmetrical.  This effect is already evident in the input RMS radial velocity variations shown in Figure~\ref{f-rms-maxjitter} and a similar effect has also found by \cite{Meunier2012} for the Sun.   The increase of the observing epochs to 100 and 200 observation has the expected impact of reducing the minimum detectable planetary mass.  A similar trend is seen in the minimum planet mass detected orbiting the K dwarf at 0.5 AU.


%

\subsubsection{I-band}


For the I-band, only the activity models Doppler image + Random Spots (DI+R), with additional plage (DI+p+R), and for 100 observations are simulated.  For the G dwarfs, the minimum detectable planet masses at 0.05 AU are the same as the V-band models (i.e. top left panel of Figure~\ref{f-summaryplots-Vband}) for the Doppler image + Random spots model (DI+R) and for the model with additional plage (DI+pl+R).  Increasing the number of observational epochs to 100 has the impact of decreasing the minimum detectable planet mass in the same way as for the V-band simulations for stellar \vsinis values $\leq$ 20 km s$^{-1}$ while also now being able to detect 5 M$_{\mathrm{J}}$ planets at \vsinis values of 50 km s$^{-1}$ and 100 km s$^{-1}$.  The wavelength dependence between V and I bands shows a similar impact on the planet detectability at 1 AU, though it is only possible to detect planets around stars with \vsinis $\leq$ 10 km s$^{-1}$ (20 km s$^{-1}$ for V-band).  This limit in \vsinis increases to 20 km s$^{-1}$ for 100 observational epochs in the I-band.


Similarly, the K dwarfs show the same global trends for the I-band models compared to the V-band models at star-planet separations of 0.05 AU.  The I-band models show more planet detections at high stellar \vsinis values, i.e. 100 km s$^{-1}$ (DI+R), though both wavelength bands show planet detection at \vsinis=100 km s$^{-1}$ for the models with additional plage (DI+pl+R) and the model using 100 observational epochs.  Increasing the star-planet separation to 0.5 AU shows a more dramatic difference with much lower mass planets being detected at all stellar \vsinis values in the I-band models.  For example, at stellar \vsinis values of 1 km s$^{-1}$,  the minimum planetary mass detected is 20 M$_{\oplus}$, compared to 50 M$_{\oplus}$ for the V-band models.  Increasing the \vsinis to 20 km s$^{-1}$ results in the detection of a 2 M$_{\mathrm{J}}$ planet  in the I-band models or 5 M$_{\oplus}$ in the V-band models and for the extreme \vsinis value of 100 km s$^{-1}$,  there is no detection in the V-band models, whereas it is possible to detect a 5 M$_{\mathrm{J}}$ planet in the I-band models.  The addition of plage has a stronger impact in the I-band models for stellar \vsinis values of 1 km s$^{-1}$ and 10 km s$^{-1}$.  The increase of observational epochs only impacts the planet detection for stellar \vsinis values $\leq$ 10 km s$^{-1}$ and 100 km s$^{-1}$,  and compared to the V-band models, the I-band consistently shows a lower minimum planetary mass that can be detected for all \vsinis values.


\subsubsection{YHK-band}


As for the I-band, only the activity models Doppler image + Random Spots (DI+R), with additional plage (DI+p+R), and for 100 observations are simulated.  The global planet detection thresholds for G dwarfs are similar to that of the I-band and V-band models for close-in planets orbiting G dwarfs, with the general trend that it is marginally easier to detect planets orbiting stars with high \vsinis values in the YHK-band.  The addition of plage results in a slight decrease in the minimum planetary mass detected for \vsinis values of 5 km s$^{-1}$, 20 km s$^{-1}$ and 50 km s$^{-1}$ which is further decreased when the number of observations is increased to 100. 

With an increased number of observational epochs, the most notable differences between I and YHK-band is for \vsinis values of 5 km s$^{-1}$ where a 50 M$_{\oplus}$ planet is detectable and for \vsinis $\geq$ 50 km s$^{-1}$ where a 2 M$_{\mathrm{J}}$ planet is detectable in contrast to 5 M$_{\mathrm{J}}$ for I-band.  In contrast to this are the V-band observations where there the minimum planetary mass detected for a stellar \vsinis value are significantly higher compared to the YHK-band.  For habitable zone planets orbiting a G dwarf, the same global behaviour between I-band and YHK-band is evident, though  for specific cases, e.g. for stellar \vsinis values $\leq$ 5 km s$^{-1}$,  the detection rate is greatly improved.


The K dwarfs also show the same global trends as for the I-band models for close-in planets, though with slightly lower planetary masses detected in the YHK-band models for \vsinis = 20 km s$^{-1}$.  For further out planets orbiting in the habitable zone of the K dwarf there is also the same global trend as the I-band, with the detection of planet at \vsinis values of 50 and 100 km s$^{-1}$ compared to a maximum of 20 km s$^{-1}$ for the V-band models.  There is also a slight improvement in planet detection for models with \vsinis = 1 km s$^{-1}$ and 5 km s$^{-1}$ (100 epochs).  The increase of the number of observations to 100 does not impact the detection of planets for the fastest rotating stars (i.e. 50 and 100 km s$^{-1}$ models) while for the I-band observations there is no change also for the model with \vsinis = 20 km s$^{-1}$.  For comparison the V-band models show that a 5 M$_{\mathrm{J}}$ planet can be detected around a star with a \vsinis = 50 km s$^{-1}$ in 100 epochs.

\section[]{Discussion}

To fully understand how planets form and evolve it is necessary to observe planets around young stars ($\leq$ 100 Myrs old).  Despite the increased stellar activity levels of these young stars, we show that it is possible to detect ``hot Jupiters" around young active stars. The ability to detect a planetary signature depends strongly on the star's \vsinis and the planet's mass and orbital radius.  With the exception of a few individual cases there does not seem to be a significant difference of planet detection with wavelength band.  Any differences that do exist are generally compensated for by extending the number of observational epochs to at least 100.
 

\subsection{Detection of Planets}
\subsubsection{Earth mass planets}
The detection of Super-Earth planets ($\leq$ 20 M$_{\oplus}$) is possible around slowly rotating G and K dwarfs with \vsinis values $\leq$ 2 km s$^{-1}$ and where the planet is orbiting close to the central star (i.e. 0.05 AU) and with 100 observational epochs in any of the wavelength bands.  Since most young G and K dwarfs rotate much faster than this, the detection of Super-Earth planets would seem unlikely, but not impossible. 

\subsubsection{Neptune mass planets}
Neptune-mass planets ($\sim$20 M$_{\oplus}$ to 159 M$_{\oplus}$ (0.5 M$_{\mathrm{J}}$) can be detected close-in (orbiting at 0.05 AU) for both G and K dwarfs for \vsinis values $\leq$ 20 km s$^{-1}$ with 100 epochs.  Such a rotation rate is quite possible for young active G and K dwarfs. While further out (1 AU for G dwarfs and 0.5AU for K dwarfs), the detection of a Neptune mass planet is only possible for stellar \vsinis $\leq$ 5 km s$^{-1}$ and 100 epochs, which is again is a rather low rotation rate for a young G and K dwarf.  Generally it is possible to detection close-in Neptune mass planets orbiting close-in to the central star on moderate rotators.

\subsubsection{Jupiter mass planets}
The detection of Jupiter mass planets is possible at all stellar \vsinis values with all activity models for close-in planets (0.05 AU) and 100 epochs or for far-out planets the detection of massive planets is only possible with stellar \vsinis values of $\leq$ 20 km s$^{-1}$ with 50 observational epochs (or 50 km s$^{-1}$ for the K dwarfs with 100 epochs).
We conclude from this that it is indeed possible to detect ``hot Jupiters" around young active G and K dwarfs.

\subsection{The effect of stellar \vsinis on planet detection}


Our results clearly show, for both K and G dwarfs, that there is a direct correlation between the magnitude of stellar activity jitter (from spots and plage) and the stellar \vsinis.  A doubling of the stellar $v$sin$i$ clearly has the impact of doubling the activity induced rms jitter, up to a limit of 50 km s$^{-1}$ where the activity features become more clearly resolved in the spectral profile.  The correlation of stellar $v$sin$i$  and the magnitude of stellar activity jitter is in agreement with our previous results for M dwarfs \citep{Barnes2011} and has a direct impact on the ability to detect planets.


In this work we have included a large range of \vsinis values to understand how \vsinis impacts planet detection thresholds, and to ultimately understand whether \vsinis or activity levels are the main limitation of detecting planets around active stars.  The large range of \vsinis values is supported  by observational evidence that shows when G and K dwarfs arrive on the zero-age-main-sequence they have a wide range of rotational velocities ranging from 0 to greater than 200 km s$^{-1}$.   Observations of rotational velocities in clusters \citep{Irwin2007,Marsden2009} show that the rotation rate of G dwarfs falls at a similar rate as K dwarfs up to 200 Myr.  After this age \citep{Irwin2007} show that the rotation rate of G dwarfs decreases steeply compared to the rotation rates of K dwarfs. 


To quantify the impact of stellar \vsinis on planet detection we modelled a large range of \vsinis values, with constant activity levels.  When the star's rotation rate is below its saturation velocity, the strength  of the magnetic field dynamo is related to the stars rotation rate.  For stars rotating faster than this, the strength of the dynamo that regenerates the star's magnetic field is no longer dependent on rotation.  This implies that, for the activity models with \vsinis values that are approximately below 20 km s$^{-1}$ \citep{Soderblom1993}, the simulated activity jitter is an upper limit rather than a realistic distribution.  There have been many tomographic maps of the surfaces of rapidly rotating G and K dwarfs with stellar \vsinis values of $\geq$ 20 km s$^{-1}$.  These images do not show a significant increase of spot coverage with stellar \vsinis.  

\subsection{The effect of activity on planet detection}

The large-scale distribution of starspots is reliably inferred from tomographic maps reconstructed using the Doppler imaging technique.  However, since the technique can only reconstruct dark starspots, we have assumed that regions of plage surround each spot with an radius twice that of the spot radius.  To account for the fact that Doppler imaging can only reconstruct large starspots we included an activity model with many small peppered random starspots in addition to the spots reconstructed using Doppler imaging techniques.

The activity model used has a strong impact on the ability to detect planets, with the most planets being detected with the activity model that uses only a Doppler image, while the least number of planet detections are for the activity model with Doppler image, plage, and random spots.  The addition of the random spots increases the activity jitter because of the additional spot coverage and also because there are many more spots located at  low latitudes.  However, the inclusion of plage acts to negate this effect, with models for the K dwarf showing an increase in the ability to detect small planets with the addition of stellar plage.  This is because there are more plage regions and at lower latitudes on the K dwarf (11\% plage coverage) compared to the high latitude plage regions on the G dwarf (6\% plage coverage).    The impact of the latitude of activity has also been noted by \cite{Barnes2011} where they showed that the latitude of the starspot impacts the amplitude of the resulting jitter, with equatorial spots resulting in a larger radial velocity amplitude than high latitude or polar spots.

In general, no two stars are the same in terms of activity level, which is supported by $\approx$ 20 tomographic images of the large-scale spot coverage on single Solar-type stars that have evolved onto the main-sequence \citep{Strassmeier2009}.  Among the G and early K dwarfs, the results of this work indicate that their spectral type is a secondary effect compared to the distribution and amount of stellar activity.  This is clearly shown in the left-hand panel of Figure~\ref{f-rms-maxjitter} where the RMS jitter for the DI+pl model is almost the same as for the K-dwarf spots and plage applied to the G dwarf (model DI+pl-G).

\subsection{The effect of observing strategy on planet detection}

In this work we have simulated a realistic range of observational epochs.  In general most planets are detected within 50 epochs.  However, for the stellar activity models with increased activity in the form of additional random spots, it is necessary to increase the number of observational epochs to 100 epochs or even 200 epochs for the lower mass planets.  The increase in the number of observational epochs to at least 100 epochs generally negates the impact of using an increased activity model.

For G and K dwarfs, the choice of wavelength band has less of an impact than for modelling the same effects on M dwarfs (see previous paper \cite{Barnes2011}).  This is because the simulated photopshere/spot contrast ratio is much larger (compared to the M dwarfs) which lessens the impact of the wavelength dependence of activity.  Plage is also shown not to have a strong impact with some examples in the V-band K dwarf models where the plage increases the planet detection threshold.  This results from plage negating the effect of spots in the stellar profile.

\subsection{Removal of activity signatures}

Recent studies have proposed that it is possible to remove the effects of activity by using a coherent spot model over the timescale of the observations \citep{Moulds2013}.  While tomographic imaging techniques such as Doppler imaging can resolve the large-scale spots, it cannot resolve the small scale structure that we have shown in our simulations to also be a key parameter in understanding how magnetic activity.  Additionally, it is well established that the spots on active stars evolve on short timescales ($\leq$ 1 month) which is significantly less than the baselines for planet detection.   One way to overcome this and to increase the detection of planets orbiting active stars is by simultaneous Doppler imaging studies, and the subsequent removal of the large-scale spot features from the radial velocity profile. This would require well sampled observations over the stellar rotation period.  While this would not remove all activity features, it would indeed significantly improve the ability to detect of planets around young active stars.

\subsection{Young planetary systems}

The results of this paper, that show it is possible to detect ``hot Jupiters" around young active G and K dwarfs (and Neptunes around more slow rotators), are very important for understanding the evolution of young planetary systems.  Theoretical models of planet formation and evolution indicate that the distribution of planets around very young stars could be very different to much older (Gyr) planetary systems, that are currently the focus of radial velocity studies, because this young stage of evolution is dominated by gravitational interactions and collisions.  By detecting Jupiter mass (and in some cases Neptune mass) planets, via the radial velocity method, orbiting young stars and comparing the number of planets and their locations to older planetary systems, it will be possible to obtain an important insight into how Jupiter and Neptune mass planets evolve in young planetary systems.
\section{Conclusions}

To fully understand planet formation and evolution it is necessary to observe much younger planetary systems, which are likely to show different distributions due to their increased dynamical interactions.  These stars are currently excluded from radial velocity searches as the increased magnetic activity of their host stars can lead to biased radial velocity signatures.  In this paper we have shown that, despite their increased magnetic activity, it is indeed possible to detect Jupiter mass planets around fast rotating stars, Neptune mass planets around moderate rotators and Super-Earths only around very slowly rotating stars.  We show that to detect planets around young active stars, surveys with significant numbers of epochs (of order 100) will be essential.


\section*{Acknowledgments}
SVJ acknowledges research funding by Deutsche Forschungsgemeinschaft (DFG) under grant SFB 963/1, project A 16. SVJ, JRB, DP and HJ have received support from RoPACS during this research, a Marie Curie Initial
Training Network funded by the European Commission’s Seventh Framework Programme. AR acknowledges research funding from the DFG grant DFG RE 1664/9-1.

\bibliographystyle{mn2e}
\bibliography{iau_journals,gdwarfs}

\label{lastpage}

\end{document}